\documentclass[aps,prl,reprint,superscriptaddress,showpacs]{revtex4-1}
\usepackage[mathlines]{lineno}
\usepackage{amsmath,amssymb,color}
\usepackage{graphicx}
\usepackage{natbib}

\usepackage{array}
\usepackage{multirow}
\usepackage{dcolumn}
\usepackage{color,colortbl}
\usepackage{hhline}
\definecolor{red}{RGB}{255,190,190}
\definecolor{lightblue}{RGB}{190,190,255}
\definecolor{green}{RGB}{190,255,190}
\definecolor{yellow}{RGB}{255,255,150}
\definecolor{grey}{RGB}{230,230,230}
\newcolumntype{P}[1]{>{\centering\arraybackslash}m{#1}}

\begin{document}

\title{All-dielectric nanophotonics: the quest for better materials and fabrication techniques}

\author{Denis G. Baranov$^\dag$}
\email[]{denisb@chalmers.se}
\affiliation{Moscow Institute of Physics and Technology, 9 Institutskiy per., Dolgoprudny 141700, Russia}
\affiliation{Department of Physics, Chalmers University of Technology, 412 96 Gothenburg, Sweden}

\author{Dmitry A. Zuev$^\dag$}
\affiliation{ITMO University, Saint Petersburg, Russia}

\author{Sergey~I.~Lepeshov}
\affiliation{ITMO University, Saint Petersburg, Russia}

\author{Oleg~V.~Kotov}
\affiliation{N. L. Dukhov Research Institute of Automatics, Moscow 127055, Russia}

\author{Alexander~E.~Krasnok}
\affiliation{Department of Electrical and Computer Engineering, The University of Texas at Austin, Austin, Texas 78712, USA}

\author{Andrey~B.~Evlyukhin}
\affiliation{Laser Zentrum Hannover e.V, Hannover, Germany}
\affiliation{ITMO University, Saint Petersburg, Russia}

\author{Boris~N.~Chichkov}
\affiliation{Laser Zentrum Hannover e.V, Hannover, Germany}
\affiliation{Leibniz Universit\"{a}t, Hannover, Germany}

\begin{abstract}
All-dielectric nanophotonics is an exciting and rapidly developing area of nanooptics which utilizes the resonant behavior of high-index low-loss dielectric nanoparticles for enhancing light-matter interaction on the nanoscale. When experimental implementation of a specific all-dielectric nanostructure is an issue, two crucial factors have to be in focus: the choice of a high-index material and a fabrication method. The degree to which various effects can be enhanced relies on the dielectric response of the chosen material as well as the fabrication accuracy. Here, we make an overview of available high-index materials and existing fabrication techniques for the realization of all-dielectric nanostructures. We compare performance of the chosen materials in the visible and IR spectral ranges in terms of scattering efficiencies and $Q$-factors. Various fabrication methods of all-dielectric nanostructures are further discussed, and their advantages and disadvantages are highlighted. We also present an outlook for the search of better materials with higher refractive indices and novel fabrication methods enabling low-cost manufacturing of optically resonant high-index nanoparticles. We hope that our results will be valuable for researches across the whole field of nanophotonics and particularly for the design of all-dielectric nanostructures.
\end{abstract}

\maketitle

\section{Introduction}
Plasmonics, the study of extraordinary optical properties of
metallic nanoparticles, has been the avant-garde of nanophotonics
for more than a decade. The intense research in this field has
arisen due to the ability of nanoparticles made of noble metals
(gold, silver) to enhance the electromagnetic field on the nanoscale
enabling the unprecedented opportunities for boosting various
optical effects and manipulation of electromagnetic radiation in
unusual ways~\cite{Pelton2008,Schuller2010,Giannini2011,Fan2014}.
However, high level of Joule losses associated with the free
electron response of noble metals was always a challenge limiting
the efficiency of optical devices~\cite{West2010,Khurgin2012,
Khurgin2015}. Although certain loss compensation approaches for
plasmonic nanostructures based on gain media~\cite{Gather2010,
Wuestner2010, Stockman2011, Pusch2012, Lisyansky2011, Andrianov2013}
and very special configuration of the metal conduction
band~\cite{Khurgin2010} have been suggested, their versatile
implementation has proven to be challenging. As a result, the
problem of Joule damping has motivated researches to search for
alternatives to noble metals, such as heavily doped
semiconductors~\cite{Naik2013} and polar crystals exhibiting
Reststrahlen bands~\cite{Feng2015}.

Mie resonances of \emph{high-index dielectric nanoparticles} pave
an alternative route towards the development of nanostructures with special
optical properties. Although Mie theory, which predicts the resonant behavior
of high-index subwavelength particles, exists for more than a century~\cite{Mie},
the enormous interest in optical properties of all-dielectric nanostructures
has arisen only recently with observation of low-order Mie modes in silicon colloids~\cite{Fenollosa, Xifre} and thanks to the advances in fabrication of single
dielectric nanoparticles with controlled geometry~\cite{Evlyukhin2012,Kuznetsov2012,Zywietz2014,Jahani2016, kuznetsov2016}.
Resonant behavior of high-index nanoparticles not only enables realization of low-loss
non-plasmonic metamaterials and metasurfaces~\cite{Wheeler2005,Popa2008, Schuller2007,Ginn2012}
with rich optical functionalities~\cite{Evlyukhin2010,Moitra2015, Arbabi2015, Shalaev2015, Khorasaninejad2015, Yu2015},
but also pave the way to enhanced light-matter interaction~\cite{Bonod2012, Albella2013,Baranov16,Krasnok2016}
as well as advanced linear~\cite{Krasnok2012, Garcia-Etxarri2011, Geffrin2012, Evlyukhin2014, Albella2015, Fu2013,Savelev2015, Li2015, Markovich2016}
and nonlinear~\cite{Shcherbakov2014, Makarov2015, Shcherbakov2015, Plasma} light manipulation. 

All-dielectric nanophotonics offers a variety of intriguing optical
effects and enables promising practical applications. In order to
implement these possibilities, \emph{materials with specific optical
properties} are desired, and certain \emph{fabrication methods} must
be available. Although the technologies of manufacturing of silicon
particles are well established today, silicon is not the only
candidate for high-index Mie resonators. There is a plethora of
semiconductors and polar materials exhibiting attractive
characteristics in visible and IR spectral regions in the context of
all-dielectric nanophotonics. Here, we present a comparative
analysis of available high-index materials in view of their
performance as optical nanoresonators. The resonant behavior is
analyzed in terms of linear characteristics of spherical
nanoparticles. To complement this analysis, we review the existing
fabrication methods of nanostructures from various high-index
materials. Finally, we provide motivations for the search of better
materials with higher refractive index and novel fabrication methods
enabling low-cost manufacturing of optically resonant dielectric
nanoparticles.

%%%%%%%%%%%%%%%%%%%%%%%%%%%%%%%%%%%%%%%%%%%%%%%%

\section{Overview of high-index materials}

\begin{figure}
\includegraphics[width=.9\columnwidth]{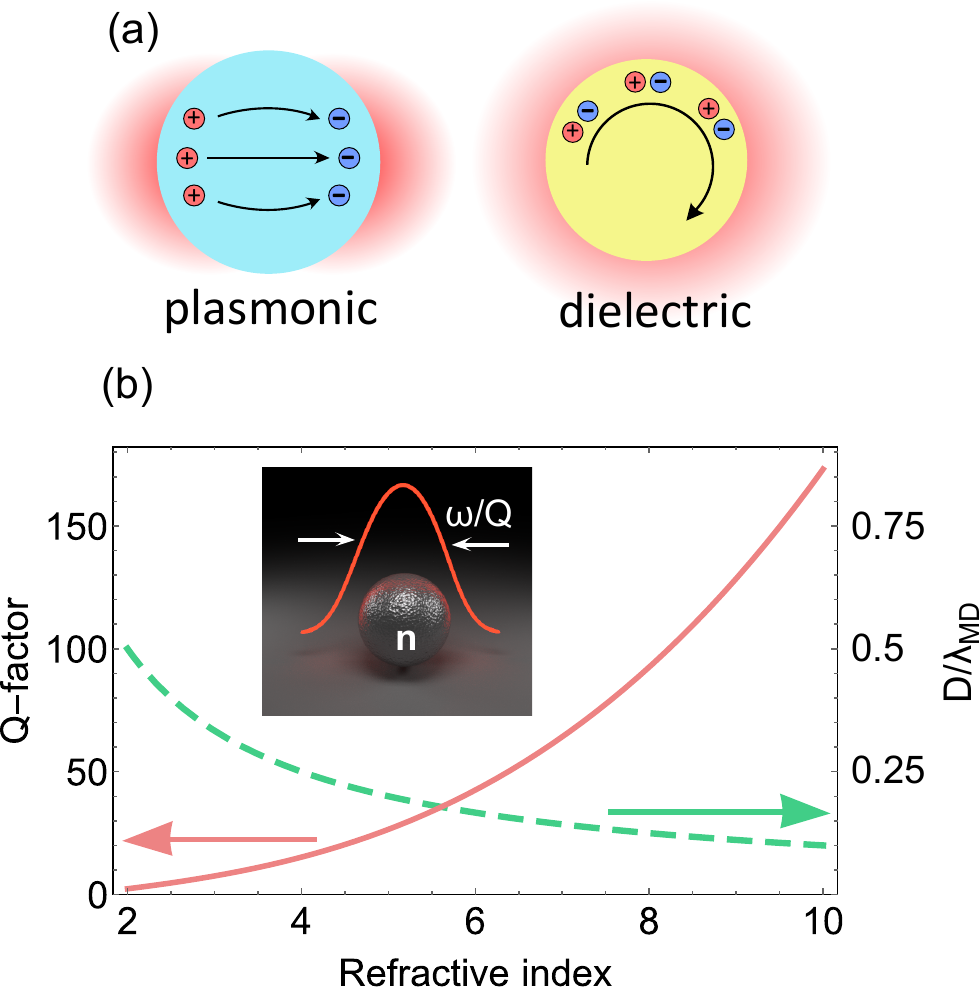}
\caption{(a) Schematic illustration of fundamental resonances and
induced currents in plasmonic and dielectric nanoparticles. Free
electrons of the metal give rise to the conduction currents in
plasmonic nanoparticles, while bounded charges in dielectrics cannot
move freely across the particle and give rise to the displacement
currents.  (b) Refractive index dependence of the $Q$-factor of the
magnetic dipole resonance of a spherical nanoparticle and the
respective dependence of the resonant nanoparticle diemater $D$. The
nanoparticle size is continuously tuned such that is satisfies the
MD resonance condition at each point.} \label{fig1}
\end{figure}

\subsection{Basics of light scattering}
%%%%%%%%%%%%   MIE
To set the stage, let us recall the basics of light scattering by
spherical particles. Electromagnetic response of a spherical
particle made of material with permittivity $\varepsilon$ of any
size is exactly described by Mie theory~\cite{Mie}. In this
framework, the electromagnetic field scattered by a sphere is
represented as a sum of electric-type (TM) and magnetic-type (TE) spherical harmonics. Amplitudes of
these harmonics are described by the Mie coefficients $a_n$ for the electric modes and $b_n$ for the magnetic modes. Each of these
amplitudes has a set of resonance frequencies, at which
electromagnetic field of the corresponding harmonic is enhanced both
inside and outside of the sphere.

Enhancement of optical phenomena by resonant nanostructures can be
described with various figures of merit. In the context of
spectroscopic applications the local electric field enhancement is
the crucial parameter which defines the efficiency of such processes
as spontaneous emission, Raman scattering, higher harmonics
generation and others~\cite{Maier06, NovotnyAOP, Agio2012}. Using
the temporal coupled mode theory, the local field enhancement
(averaged over the mode volume) may be expressed as
follows~\cite{Seok11}:
\begin{equation}
{\left( {\frac{{{E_{{\text{loc}}}}}}{{{E_{{\text{inc}}}}}}} \right)^2} \propto \gamma_{\rm rad} \frac{Q^2}{V},
\label{eq1}
\end{equation}
where $\gamma_{\rm rad}$ is the nanoantenna radiative damping rate,
$Q$ is the quality factor of the excited mode, $V$ is its mode
volume. Expression~(\ref{eq1}) suggests that high $Q$ and small $V$
are beneficial for the local field enhancement with nanoantennas.

In contrast to plasmonics, where the Joule losses represent the
dominant dissipation channel, the $Q$-factor of high-index Mie
resonators is mainly limited by the radiation damping. In the
context of metasurfaces, which manipulate the characteristics of
propagating fields, radiation should not be treated as a loss. For
this reason, a relevant figure of merit is the nanoantenna radiation
efficiency $\eta_{\rm rad}$:
\begin{equation}
\eta_{\rm rad}=\frac{\sigma_{\rm scat}}   { \sigma_{\rm scat}+\sigma_{\rm abs}}
\label{eq2}
\end{equation}
with $\sigma_{\rm scat}$ and $\sigma_{\rm abs}$ being the scattering
and absorption cross-sections, respectively. Value of $\eta_{\rm
rad}$ close to 1 indicates that almost all incident light is
re-radiated without being absopbed by nanoantennas constituing a
metasurface. At the same time, a smaller nanoantenna size is also
desired since it would allow smaller distances between neighboring
nanoantennas reducing the spatial dispersion effect.

%plasmonic vs dielectric
The extraordinary enhancement of electric field by metallic
nanoparticles relies on free electron response, Fig.~\ref{fig1}(a).
When the frequency of incident light matches that of the free
electron oscillations inside a sphere, strong electric field is
produced in the vicinity of the particle. These oscillations,
however, are accompanied by significant optical loss arising from
both intraband and interband transitions and eventually lead to
heating of the resonator~\cite{ Khurgin2015}. In contrast, optical
resonances of high-index nanoparticles originate from the
\emph{displacement currents} due to oscillations of bounded
electrons, Fig.~\ref{fig1}(a). These currents are free of Ohmic
damping, what allows to reduce non-radiative losses and heating of
an optical cavity.

%MD
Of special interest to the field of nanophotonics is the
\emph{magnetic dipole (MD) resonance} of high-index nanoparticles -
the fundamental magnetic dipolar mode of a dielectric sphere. At a fixed
nanosphere diameter, the MD resonance occurs at the smallest frequency of the incident wave compared to other resonances~\cite{Evlyukhin2010,Evlyukhin2012}. Under the
resonance condition, electric fields are anti-parallel at the
opposite boundaries of the sphere, which gives rise to strong
coupling to circulating displacement currents characteristic for a
magnetic dipole mode~\cite{Zywietz2014}.

The spectral position of the MD resonance of a spherical particle is
approximately defined by ${\lambda _{{\text{MD}}}} \approx nD$ with
$n$ being the refractive index of the sphere and $D$ its diameter.
Therefore, larger refractive indices are desirable from the point of
view of device miniaturization. The field enhancement and $Q$-factor
of Mie resonances also benefit from the large particle refractive
index (see Fig.~\ref{fig1}(b)) -- this may be intuitively understood as
the result of smaller radiation leakage from nanoparticles with
larger refractive index contrast.

%%%%%%%%%%%%%%%%%    semicond
\subsection{High refractive index materials}
\begin{figure}[t]
\includegraphics[width=.9\columnwidth]{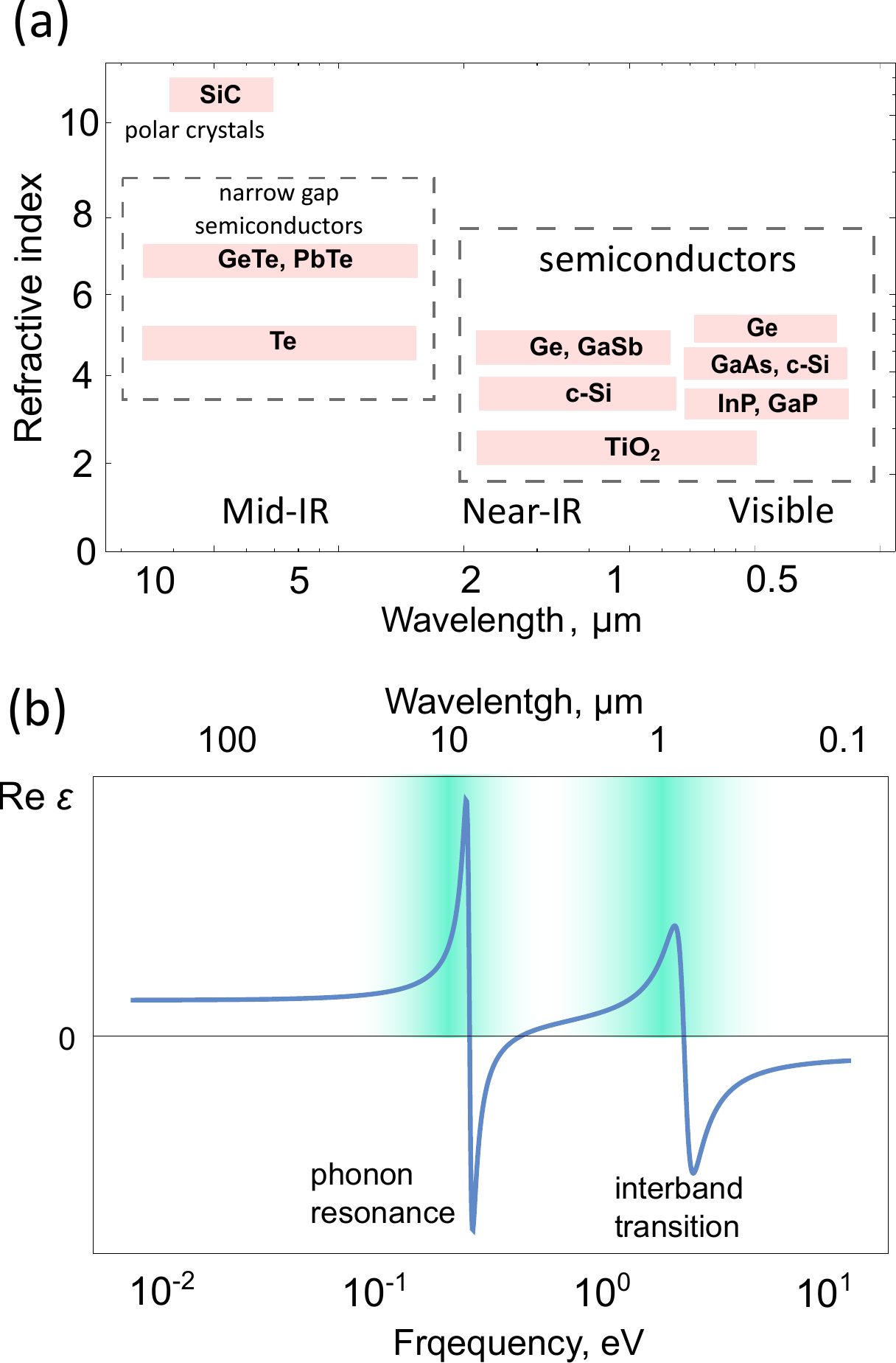}
\caption{(a) Refractive indices of materials available for
dielectric nanophotonics from visible to mid-IR spectral ranges. (b)
A typical dielectric response of a high-index material exhibiting a
series of resonances. Shaded area depicts two high-index plateau
related to material resonances.} \label{index}
\end{figure}

The above considerations clearly pose the \emph{quest for materials}
with large refractive indices. In the visible and near-IR spectral
ranges, the highest known permittivities are demonstrated by
semiconductors such as Si, Ge, GaSb and others, see
Fig.~\ref{index}(a). In the neighboring mid-IR range, which is also
of great interest to the optoelectronic technology, narrow-band
semiconductors and polar crystals demonstrate very attractive
characteristics, and have been implemented for the design of
all-dielectric metamaterials consisting of resonant
Te~\cite{Ginn2012} and SiC~\cite{Schuller2007} elements.

%resonances and absorption
The origin of relatively high refractive index of these materials
can be understood from the picture of their electronic response, see
Fig.~\ref{index}(b). While in the low frequency regime the response
is purely dielectric (assuming that we are dealing with undoped
materials), in the visible and IR spectral ranges these materials
exhibit a series of resonances. Due to the coupling of light to
these resonances of the medium, the regions of increased refractive
index appear in the spectrum. The low frequency phonon-polariton
resonance occurs due to coupling of light with optical phonons of
the crystal lattice of polar crystals~\cite{Klingshirn, Cardona}.
The higher frequency exciton-polariton resonance originates from
interband transitions in semiconductors and forms a pronounced
plateau in their dielectric function.

The high refractive index of these materials comes at a cost of
increased absorption. According to the Kramers-Kronig
relations~\cite{Nussenzveig}, any dispersion of permittivity is
related to dissipation. As it was pointed above, the high-index
regions of semiconductors and polar crystals stem from their exciton
and phonon resonances which bring optical absorption. This
fundamental trade-off between absorption and high refractive index
eventually sets the limit to all-dielectric resonator performance.

There is a significant difference between the behavior of refractive
indices of semiconductors and polar crystals. In semiconductors, the
high refractive index at below-band gap frequencies originates from
a continuum of interband transitions~\cite{Cardona}, what allows one
to have high index and relatively low absorption at the same time.
In polar crystals, high index is related to a single phonon
resonance, so that it is accompanied by large absorption
coefficient~\cite{CaldwellNanoph}.

\begin{table*}
\renewcommand{\arraystretch}{1.4}
  \begin{tabular}{| P{2.0cm} | P{2.8cm} |P{2.2cm} | P{2.2cm} | P{2.1cm} | P{1.7cm} | P{1.5cm} |}
\hline
    Material & Spectral range, $\mu m$ & Refractive index, n & Extinction coefficient, k & Bandgap type & Bandgap energy, eV & Reference \\ \hline
    \multirow{2}{*}{c-Si} & 0.50 -- 1.45  & 4.293 -- 3.486 & 0.045 -- 0.001 & \multirow{2}{*}{indirect} & \multirow{2}{*}{1.12} & \citenum{Green1995} \\ \hhline{~---~~-}
    & 1.45 -- 2.40 & 3.484 -- 3.437 & $\sim$0 &  &  & \citenum{Li1980}  \\ \hline
    a-Si & 0.50 -- 1.00 & 4.47 -- 3.61 & 1.12 -- 0.01 & indirect & 1.50 & \citenum{Pierce1972}  \\ \hline
    GaAs & 0.50 -- 0.80 & 4.037 -- 3.679 & 0.376 -- 0.089 & direct & 1.46 & \citenum{Jellison1992}  \\ \hline
    GaP & 0.50 -- 0.80 & 3.590 -- 3.197 & $\sim$0 & indirect & 2.26 & \citenum{Jellison1992}  \\ \hline
    InP & 0.50 -- 0.80 & 3.456 -- 3.818 & 0.203 -- 0.511 & direct & 1.27 & \citenum{Aspnes1983}  \\ \hline
    TiO$_{\rm 2}$ & 0.50 -- 1.00 & 2.715 -- 2.483  & $\sim$0 & indirect & 3.05 & \citenum{Devore1951}  \\ \hline
    \multirow{4}{*}{Ge} & 0.50 -- 0.60  & 4.460 -- 5.811 & 2.366 -- 1.389 & \multirow{4}{*}{indirect} & \multirow{4}{*}{0.67} & \multirow{2}{*}{ \citenum{Jellison1992}  } \\ \hhline{~---~~~}
    & 0.60 -- 0.80 & 5.811 -- 4.699 & 1.389 -- 0.3 &  &  &  \\ \hhline{~---~~-}
    & 0.80 -- 1.90 & 4.684 -- 4.129 & 0.3 -- 0.001 &  &  & \citenum{Palik} \\ \hhline{~---~~-}
    & 1.90 -- 2.40 & 4.111 -- 4.069 & $\sim$0 &  &  & \citenum{Li1980} \\ \hline
    GaSb & 1.00 -- 2.40 & 4.140 -- 3.846 &  0.225 -- 0.001 & direct & 0.69 & \citenum{Ferrini1998}  \\ \hline
    Te & 4.00 -- 14.0 & 4.929 -- 4.785 &  $\sim$0 & indirect & 0.34 & \citenum{Caldwell1959}  \\ \hline
    PbTe & 4.10 -- 12.5 & 5.975 -- 5.609  &  $\sim$0 & direct & 0.31 & \citenum{Weiting1990}  \\ \hline
    GeTe & 6.20 -- 11.8 & 7.3 -- 7.278  &  $\sim$0 & direct & 0.2 & \citenum{Okoye2002}  \\ \hline
    SiC & 11.0 -- 15.0 & $\sim$ 20 &  $\sim$ 15 & - & - & \citenum{Larruquert2011}  \\ \hline
  \end{tabular}
  \caption{Optical properties of high index materials.}
  \label{tabl}
\end{table*}

The value of refractive index of a semiconductor is closely related
to its electronic band gap. Generally, the \emph{electrostatic}
refractive index of a semiconductor decreases with the increasing
energy gap. This correlation can be understood from the following
argument, suggested by Moss~\cite{Moss}. He considered electrons in
semiconductor as if they were bound to a Hydrogen atom. The energy
needed to ionize the atom and to raise an electron to the conduction
band scales as $E_g \sim 1/\varepsilon^2$, where $\varepsilon$ is
the background permittivity of the semiconductor. This results in a
very simple approximation for the static refractive index known as
the Moss formula:
\begin{equation}
n^4 E_g=95~\text{eV}.
\end{equation}
There are several other models leading to similar
dependencies~\cite{Ravindra, Tripathy}. Although being valid for
static permittivity, this approximation qualitatively reproduces
behavior of refractive indices at optical and IR wavelengths,
summarized in Fig.~\ref{index}(a): overall, narrow gap
semiconductors such as PbTe, GeTe demonstrate higher refractive
indices than those with energy gap lying in the visible. The
detailed information on optical properties of chosen materials
including values of the extinction coefficients and electronic band
gap as well as the sources from which the data on permittivity
dispersion is presented in Table~\ref{tabl}.

Semiconductor materials may have either direct or indirect band gap,
what has a profound implication on the absorbing properties. When
light passes through a direct bandgap semiconductor, a photon can be
absorbed to the conduction band. At the same time, when light
travels through an indirect band gap medium, due to large mismatch
between the electron and photon wavevectors, such process does not
occur what results in reduced optical absorption. Nevertheless, even
direct band gap materials may demonstrate good performance if the
operation frequency lies below the band gap, such as the case of GaSb in the mid-IR range (see below).

%%%%%%%%%%%%%%%%%%%%%%%%%%%%%%%%%%%%%%%%                    Q factors            %%%%%%%%%%%%%%%%%%%%%%
\subsection{Comparative analysis of materials}

\begin{figure*}
\includegraphics[width=2\columnwidth]{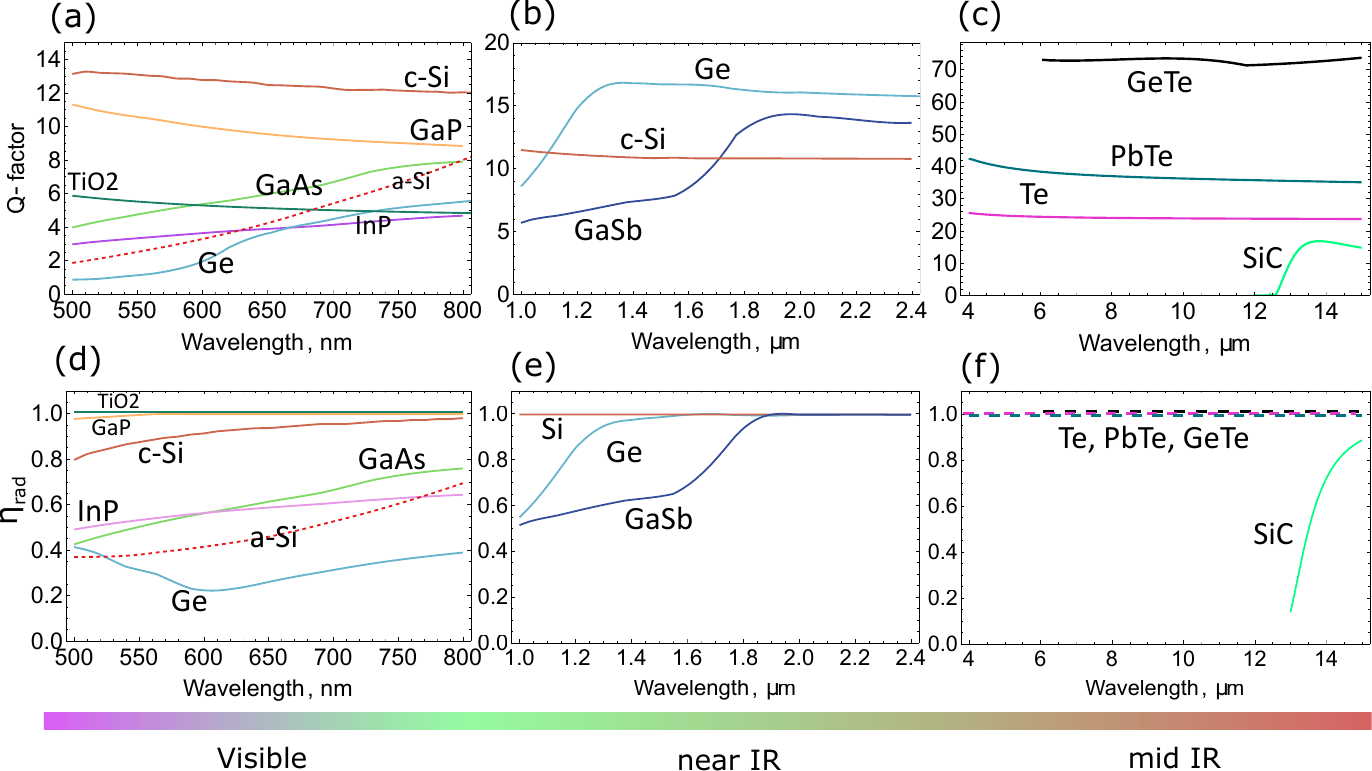}
\caption{Quantitative comparison of the available high-index
materials for all-dielectric nanophotonics. (a)--(c) $Q$-factors of
the magnetic dipole (MD) resonance of spherical nanoparticles made
of various high-index materials as a function of wavelength in the
visible (a), near-IR (b) and mid-IR (c) regions. (d)--(f) The
antenna radiation efficiency $\eta_{\rm rad}$ at the MD resonance
for the same scope of materials.} \label{fig2}
\end{figure*}

\begin{figure*}
\includegraphics[width=2\columnwidth]{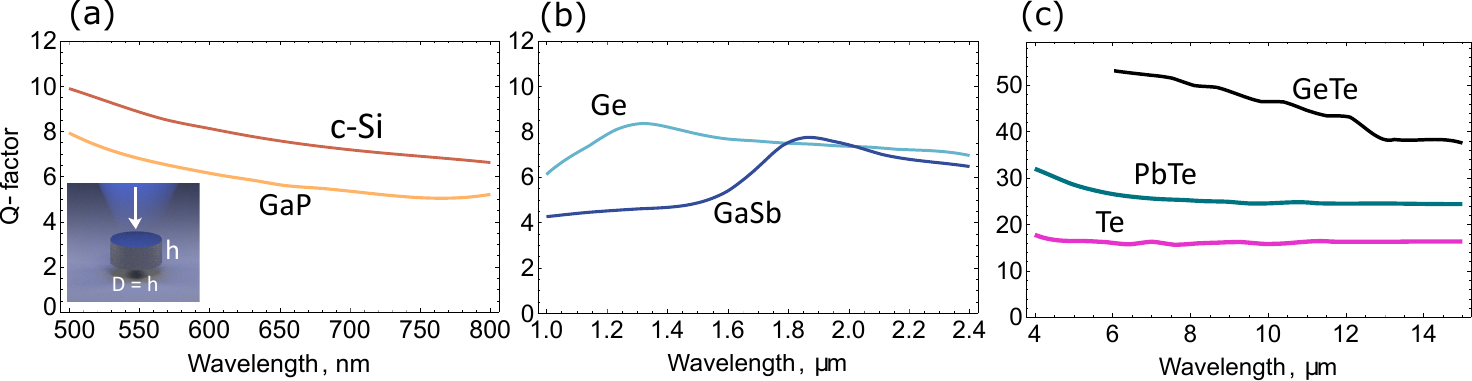}
\caption{$Q$-factors of the MD resonance of nanodisks with equal height and diameter. The nanodisk sizes were continously tuned such that is satisfies the
MD resonance condition at each wavelength.}
\label{disks}
\end{figure*}

Let us now illustrate performance of the available high-index
materials for the case of a spherical particle since it allows exact
analytical solution. First, we present in Figs.~\ref{fig2}(a)--(c)
the $Q$-factor of the MD resonance of a spherical particle as a
function of wavelength. In order to obtain this dependence, we
continuously tune the nanoparticle size such that is satisfies the
MD resonance condition at each wavelength. The $Q$-factor was
obtained by calculating the full-width of the MD resonance
scattering cross-section. The results show that across the whole
visible spectral range among all semiconductors, c-Si demonstrates
the highest $Q$-factor of around 13 with the closest being GaP with
$Q$-factor of only around 10. Good performance of c-Si and GaP is
caused by the fact that these two materials are indirect band gap
semiconductors, what results in the reduced optical absorption and
therefore higher $Q$-factors. In the near-IR, Ge takes the pedestal
with $Q$-factor slightly above 15 at the most interesting wavelength
of 1.55 $\mu$m. Notably, GaSb demonstartes good values of the $Q$-factor at wavelengths above 1.8 $\mu$m despite being a direct semiconductor.
In the mid-IR, the situation is strikingly
different: particles of narrow band gap semiconductors PbTe and GeTe
demonstrate $Q$-factor well above 40 and 70, respectively. SiC, on the other hand,
despite its huge refractive index exceeding 10, shows very low
quality factor. Such poor optical response of SiC nanoparticles
stems from narrow phonon resonance of the polar crystal associated with large absorption, as argued above.

The results presented in Figs.~\ref{fig2}(d)--(f) indicate that the
radiation efficiencies of semiconductor nanoparticles $\eta_{\rm
rad}$ calculated with Eq.~\ref{eq2} generally increase with the
increasing wavelength. Interestingly, in the visible range the
highest radiation efficiency is demonstrated by TiO$_2$ and GaP
nanoparticles due to the widest electronic bandgap and thus the
smallest absorption among the studied materials, see
Table~\ref{tabl}. In the near-IR, the efficiency of Ge and Si
antennas is nearly one at wavelengths longer than 1.5 $\mu$m. In the
mid-IR, the efficiencies of Te, PbTe and GeTe particles are equal to
1, while SiC particles show very poor performance with low radiation
efficiency and $Q$-factor owing to the sharp phonon resonance.

Although the spherical geometry of a particle is attractive from the
theoretical standpoint, as it allows exact analytical description of
light scattering, fabrication of spherical particles is not always
an option. On the other hand, Mie resonances occur for a wide range
of particle shapes, including nanodisks which may be easily
fabricated by the standard nanolithography techniques (see Section
III A) for any material considered in this study. For this reason,
we performed the same analysis of $Q$-factors for nanodisks made of
different materials. The results shown in Fig.~\ref{disks} demonstrate that behavior of $Q$-factors is overall similar to that observed for spherical particles.
To conclude this section, we finally note that the observed $Q$-factor values do not set the upper limit for high-index nanostructures, and larger $Q$-factors can be obtained in clusters of nanoparticles, such as dimers and metasurafces~~\cite{ZywietzACSPhoton2015,Shalaev2015,campione2016broken}.

%%%%%%%%%%%%%%%%%%%%%%%%%%%%%%%%%%%%%%%%%%%%%%%%%%%%%%%%%%%%%%%%%%%%%%%%%%%%%%%%%%
\section{Overview of fabrication techniques}

\begin{table*}
\renewcommand{\arraystretch}{1.4}
\centering
  \begin{tabular}{| P{4.0cm} | P{5.5cm} | P{5.5cm} |}
\hline
    \rowcolor{grey}Method & Advantages & Disadvantages \\ \hline
    \rowcolor{red} Lithography~\cite{Fenollosa, ShiAdvMat2012, Polman, Seo, Feng, StaudeACSNano2013, SpinelliNC2012, Novotny2013, Sinclair2013, Polman2013,StaudeReview}  & \begin{flushleft}  - high resolution (below 10 nm) \linebreak - high repeatability \linebreak - complicated nanostructures can be fabricated \end{flushleft} & \begin{flushleft} - non-single step process \linebreak - spherical shape is not accessible in nanolithography methods  \linebreak - complicated equipment \end{flushleft}  \\   \hline
    \rowcolor{green} Thermal dewetting~\cite{AbbarchiACSNano2014, AbbarchiNanoscale, Anderson, Lagally} & \begin{flushleft} - high-productivity \linebreak - simplicity \end{flushleft} & \begin{flushleft} - precision control of size and location of the nanoparticles can be achieved only via additional methods \end{flushleft}  \\ \hline
    \rowcolor{green} Chemical methods~\cite{Proust, ShiNC2013, Brongersma, Schuller2007} & \begin{flushleft} - high-productivity \linebreak - nanoparticles can be fabricated in colloid immediately \end{flushleft} & \begin{flushleft} - chemical waste \linebreak - necessity of nanoparticles additional ordering \linebreak - contamination of nanoparticles during fabrication is possible \end{flushleft} \\ \hline
     \rowcolor{lightblue} Laser-assisted methods~\cite{Kuznetsov2012, Fu2013, Lewi2015, Okamoto, Evlyukhin2012, ZywietzAPA2014, Zywietz2014, ZywietzACSPhoton2015} & \begin{flushleft} - single step process \linebreak - simplicity \linebreak - high repeatability \linebreak - lack of harmful chemical waste \linebreak - amorphous or crystalline nanoparticles can be produced \end{flushleft} & \begin{flushleft} - need in a film preparation for the process \linebreak - high reliability of the mechanic systems in experimental setup \linebreak - high demands to focusing and laser beam shape \end{flushleft} \\ \hline
  \end{tabular}
  \caption{Summary of the available methods for fabrication of all-dielectric nanostructures.}
  \label{methods}
\end{table*}

Bearing in mind the above considerations about high-index materials, we
now may proceed to a brief review of the existing fabrication
techniques. The rapid progress of nanotechnology has enabled
tremendous development of methods for semiconductor nanoparticle
fabrication. The most illustrative example is presented by silicon, since it is
the most frequently used high-index material in the visible and IR
ranges owing to its relatively low cost and low imaginary part of
the refractive index. Therefore, it is not surprising, that methods
of fabrication of silicon nanostructures have been first
historically developed. Initially, different methods for fabrication
of optically small and non-resonant silicon nanoparticles were
developed basically for biological imaging and drug delivery
(Ref.~\citenum{Chu}), including methods of mechanical
milling~\cite{Timoshenko}, pulsed laser ablation in
liquid~\cite{KabashinJMCB}, electrochemical etching~\cite{Sailor},
gas phase synthesis~\cite{Peukert}, etc.

In the context of nanophotonic applications, nanoparticles
with the magnetic and electric resonances located in the visible or
near-infrared spectral regions are desired. This requirement sets up
high demands not only to the nanoparticle size, but also to
repeatability and throughput of fabrication methods, precise
control of nanoparticle geometry, simplicity of the fabrication
procedure (quantity of steps involved in the fabrication process),
and manipulation of nanoparticle space arrangement.
Therefore, the development of fabrication technologies of
Mie-resonant high-index nanoparticles has recently been initiated,
that resulted in emergence of various techniques, roughly summarized
in four groups in the Table~\ref{methods}. This table demonstrates
that all techniques have inherent advantages and disadvantages.
Thus, the choice of a fabrication method of high-index nanoparticles
is determined by the objectives of the investigation, e.g. in case
of high demands to chemical purity, laser-assisted methods are
preferable, in contrast, if high-productivity is a cornerstone,
chemical methods will be considered as more preferable ones. To
better understand applicability of these methods for a special case,
they should be carefully considered on the example of fabricated
high-index nanostructures.

\subsection{Lithography}
\subsubsection{Single nanostructures}
\begin{figure}[t]
\includegraphics[width=.9\columnwidth]{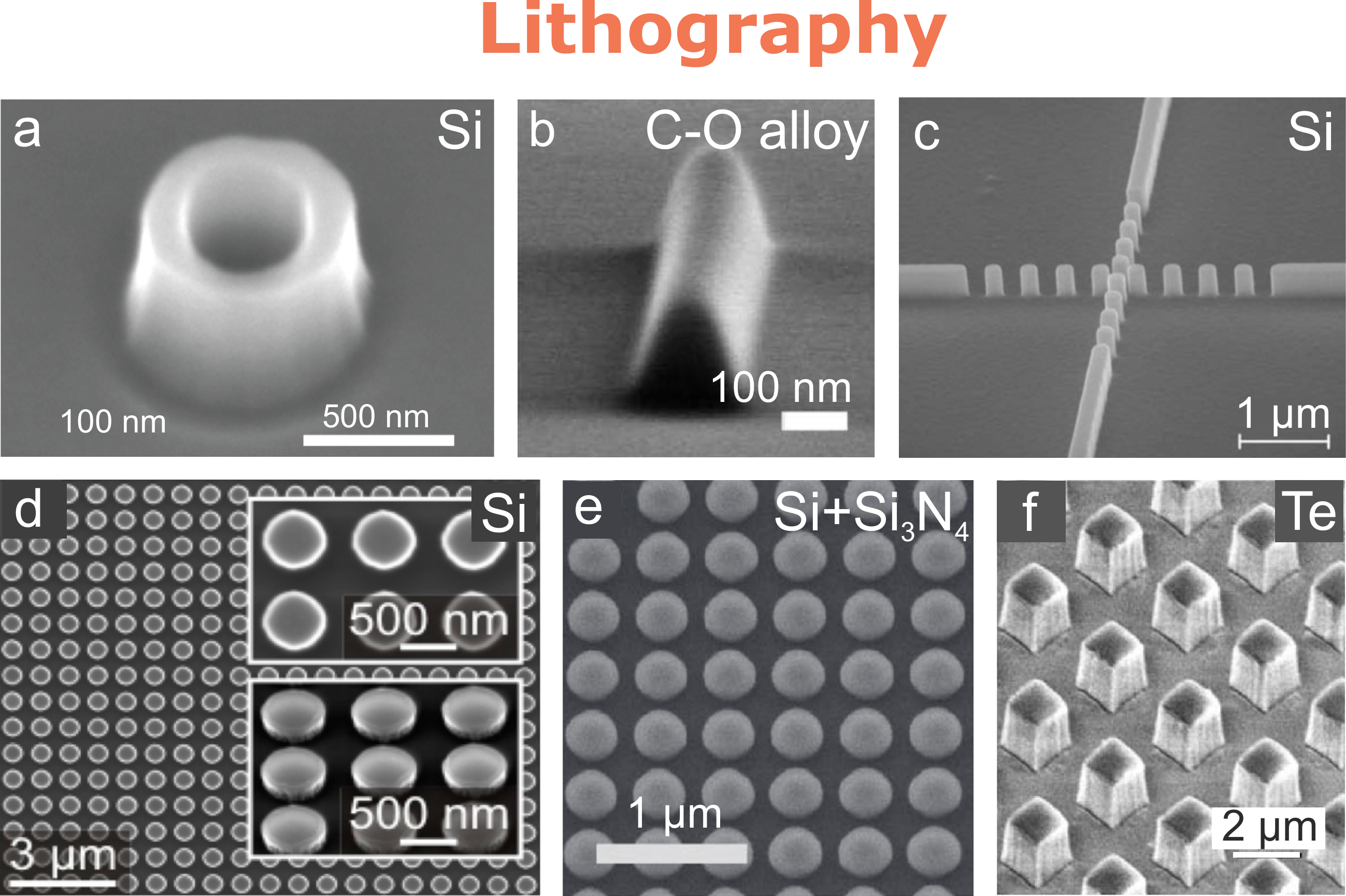}
\caption{(a) Scanning electron microscopy (SEM) image of hollow Si cylinder. Reprinted with permission from Ref.~\citenum{Polman}. (b) SEM image of carbonaceous dielectric nanorod antenna fabricated by electron beam-induced deposition. Reprinted with permission from Ref.~\citenum{Seo}. (c) Periodic dielectric waveguide crossing. Reprinted with permission from Ref.~\citenum{Feng}. (d) Silicon nanoparticles obtained by means of reactive-ion-etching through a mask. Reprinted with permission from Ref.\citenum{StaudeACSNano2013}. (e) Si nanoparticles with additionally deposited Si$_3$N$_4$ thin film, reprinted with permission from Ref~\citenum{SpinelliNC2012}. (f) High-index cubic tellurium resonators, reprinted with permission from Ref.~\citenum{Ginn2012}}.
\label{fig:Fabr_1}
\end{figure}

The most straightforward methods for fabrication of nanostructures
involve lithography, since it provides high repeatability alongside
with an opportunity to fabricate nanostructures of complicated
shapes via combination of lithographic processes. The conventional
lithographic  methods have been successfully applied for the
fabrication of single nanostructures. For example, silicon
nanoresonators consisting of hollow nanocylinders with an outer
diameter of 108$-$251~nm and a gap size $>$20~nm were fabricated by
the combination of electron beam lithography with reactive ion
etching, Fig.~\ref{fig:Fabr_1}(a)~\cite{Polman}. The geometry of
these structures (the outer diameter, height and wall thickness) can
be varied to control the resonant wavelength and relative spectral
spacing of the modes.

In Ref.~\citenum{Seo}, a single dielectric nano-rod antenna (see
Fig.~\ref{fig:Fabr_1}(b)) consisted of amorphous alloys of C and O
was fabricated via electron beam-induced deposition. The control of
nanoantenna geometry via fabrication process was applied for
resonant light scattering control over the whole visible wavelength
range. In Ref.~\citenum{Feng}, the intersection region of two
silicon waveguides was replaced by an array of silicon cylinders,
Fig.~\ref{fig:Fabr_1}(c). The proposed method of high-index
nanostructure fabrication is very promising for optical cross
connects because the cylinder structure is compatible with the
existing lithography processes, can be fabricated synchronously with
waveguide components as well as demonstrates high transmission and
negligible cross talk over a broad bandwidth.

\subsubsection{Nanostructure arrays}

The standard micro/nanofabrication processing techniques are proven
to be effective means of fabrication of large scale nanostructure
arrays. The fabrication of large-scale arrays is very important for
the creation of high-index metasurfaces setting higher demands to
the precision and reproducibility of nanostructures. The
controllable fabrication of silicon nanoparticle arrays was achieved
by a multi-step method, including electron-beam lithography on
silicon-on-insulator wafers (formation of mask from resist) and
reactive-ion-etching process with following removal of remaining
electron-beam resist mask. This technology allows forming arrays of
silicon nanodisks, Fig.~\ref{fig:Fabr_1}(d)
\cite{StaudeACSNano2013}, in which Mie-resonances can be precisely
tuned by varying basic geometrical parameters (diameter and height).
More complicated nanostructures representing the combination of Si
nanoparticles, produced on the silicon wafers by substrate conformal
soft-imprint lithography in combination with reactive ion etching,
and a Si$_3$N$_4$ coating, Ref.~\citenum{SpinelliNC2012} (see
Fig.~\ref{fig:Fabr_1}(e)) are demonstrated for the average
reflectivity reduction up to the values lower than 3\% over the wide
spectral range of 450$-$900~nm.

Multi-step lithography methods have been also implemented for the
fabrication of high-index tellurium dielectric resonators for
all-dielectric metamaterial, see
Fig.~\ref{fig:Fabr_1}(f)~\cite{Ginn2012}. For these experiments,
BaF$_{2}$ optical flat substrate was applied for deposition of
1.7~$\mu$m Te thick film by electron-beam evaporation. Then, this
film was patterned using electron-beam lithography and a reactive
ion etching process to fabricate uniform structure. Zero
backscattering and significant light dispersion in forward direction
were demonstrated in GaAs nanoparticles, fabricated with molecular
beam epitaxy, lift-off procedure, reactive ion and plasma
etching~\cite{Novotny2013}. The Te dielectric metamaterial structure
was also made in Ref.~\citenum{Sinclair2013} using multi-cycle
deposition-etching-liftoff technique. In Ref.~\citenum{Polman2013},
an array of TiO$_{2}$ nanostructures on a Al$_{2}$O$_{3}$-passivated
Si wafer was produced by a combination of standard lithography
methods: substrate-conformal imprint lithography, reactive ion
etching and plasma assisted atomic layer deposition (ALD). These
procedures made possible fabrication of nano-patterned dielectric
coating for crystalline Si solar cells possessing anti-reflection
properties due to preferential forward scattering of light via
proper engineering of Mie resonances of TiO$_{2}$ nanocylinders. The combination of e-beam lithography and dielectric reactive ion etching was also successfully implemented to metasurface demonstrating Q-factor of ~350 (for Si-based structure) and Q-factor of ~600 (for GaAs-based structure) in the near-infrared (~1 $\mu$m)~\cite{campione2016broken}
 
Another approach to fabrication of planar optical metasurfaces based on high-quality TiO$_{2}$ grown by ALD has been demonstrated in Refs.~\citenum{Khorasaninejad2015,devlin2016broadband} (see also Ref.~\citenum{genevet2017recent} for a review of planar optical metasurfaces). Here, ALD provides precise control of film thickness (up to a monolayer) and the material phase, making possible creation of high aspect-ratio nanostructures with no losses in the visible region. The proposed approach can be applied to any type of metasurfaces.

Thus, lithographic methods remain more reliable for high-index
nanostructure fabrication combining three important qualities: high
reproducibility for the nanostructure arrays, possibility of
fabrication of nanostructures with complex shape, and high
resolution. More information on lithographic fabrication of dielectric
nanostructures can be found in Ref.~\citenum{StaudeReview}.
It should be mentioned, that in spite of these advantages, the methods can hardly be
applied for the creation of spherical nanoparticles that represent
interest not only for fundamental but also for applied sciences.
Moreover, multiple steps, that are often required for high-index
nanostructure fabrication, as well as, complexity and high cost of
equipment motivate researchers to look for novel fabrication
methods.

\subsection{Chemical methods}

Chemical methods are promising in terms of high-throughput
fabrication of nanoparticles. These methods also offer flexibility
in synthesis of materials, relative simplicity, and compatibility
with other methods of solid material synthesis. One of the most
widespread methods is the  chemical vapor deposition technique. This
process can be applied as an effective method for the fabrication of
silicon nanoparticles with different sizes: e.g. fabrication of
silicon nanoparticles can be carried out by decomposition of
disilane gas (Si$_2$H$_6$) at high temperatures into solid silicon
and hydrogen by the following chemical reaction: Si$_2$H$_6
\rightarrow$ 2Si(s) + 3H$_2$(g). As a result, spherical polydesperse
silicon particles with diameters from 0.5 to 5 $\mu$m
(Ref.~\citenum{Fenollosa}) and about $300-500$~nm
(Ref.~\citenum{ShiAdvMat2012}) were produced by this method. The
decomposition temperature of the precursor as well as annealing
treatment can be chosen to fabricate amorphous or polycrystalline
particles with a crystallite size of about 3~nm. It is interesting, that appearance of unwanted silicon dust in the decomposition of silicon precursors is mentioned in Ref.~\citenum{o1990handbook}. In fact, this dust is composed of silicon nano and microspheres, as it is demonstrated in Refs.~\citenum{Fenollosa,ShiAdvMat2012}.

\begin{figure}[!t] \centering
\includegraphics[width=.8\columnwidth]{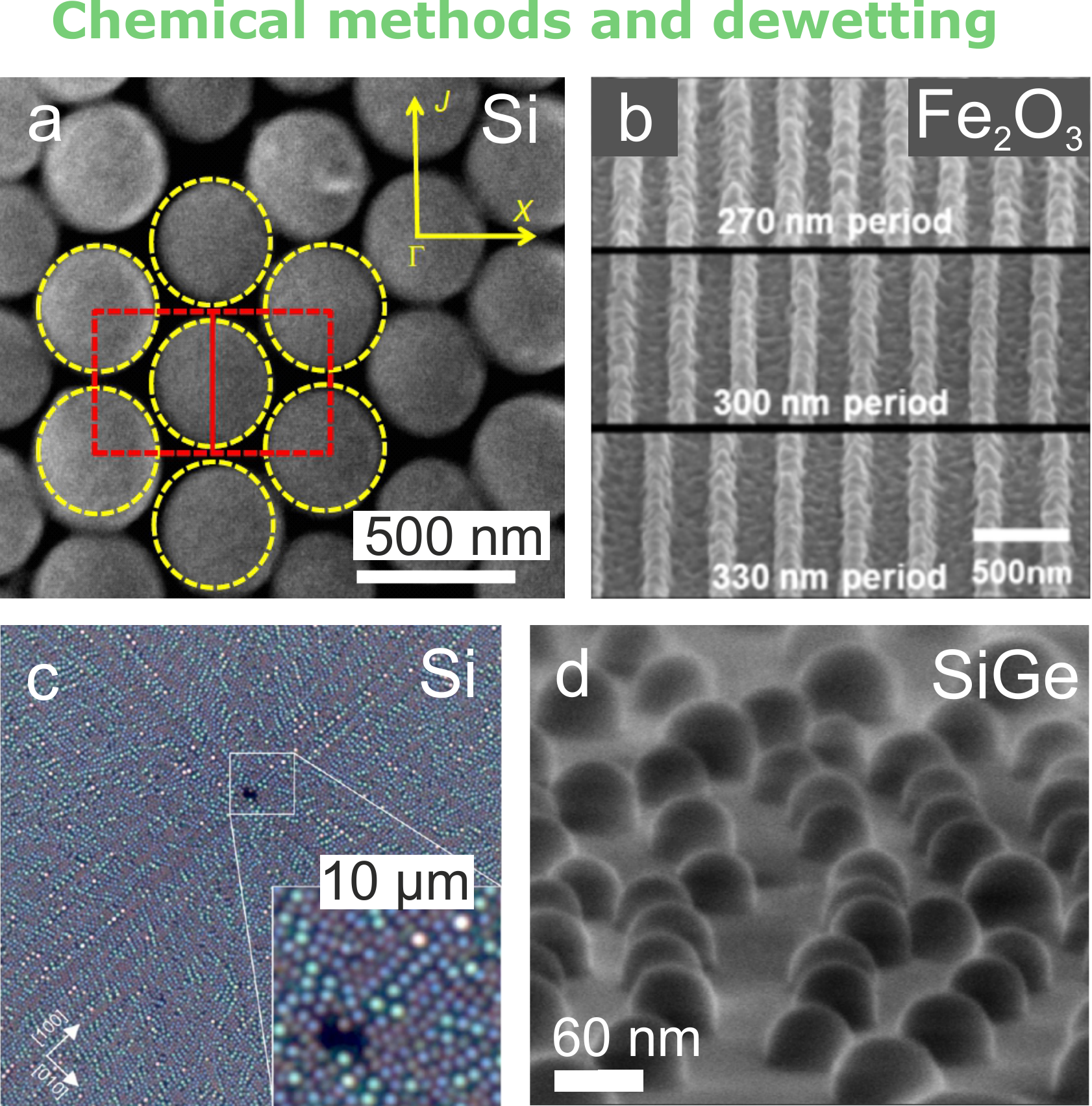}
\caption{(a) self-aligned
silicon nanoparticles obtained by chemical
deposition, reprinted with permission from Ref.~\citenum{ShiNC2013}. (b) Periodic Fe$_2$O$_3$ nanobeam-array
with different periods fabricated via combination of lithography
(top image) and oxidation in air (bottom), reprinted with permission from Ref.~\citenum{Brongersma}. (c) Dark-field optical image of silicon nanoparticles obtained by thin
film dewetting, reprinted with permission from Ref.~\citenum{AbbarchiACSNano2014}. (d) SiGe nanoparticles received after thermal dewetting, reprinted with permission from Ref.~\citenum{Lagally}.}
\label{fig:Fabr_2}
\end{figure}

The crystalline silicon Mie resonators can be also created via
alkaline chemical etching combined with electronic
lithography~\cite{Proust}. This method eliminates reactive ion
etching and can be applied for the fabrication of both silicon
nanoresonators and oligomers. Further, fabrication of a
monodispersed silicon colloid was achieved via decomposition of
trisilane (Si$_3$H$_8$) in supercritical n-hexane at high
temperature~\cite{ShiNC2013}. In this method, the particles size can
be controlled by changing the trisilane concentration and reaction
temperature. With this method a plenty of similar silicon
nanoparticles with size dispersion in the range of several percents,
which can be ordered into hexagonal lattice by self-assembly,
Fig.~\ref{fig:Fabr_2}(a), can be obtained. The main disadvantage of
this method is the porosity and high hydrogen content of
nanoparticles, as well as, the necessity of additional patterning
methods to fabricate functional structures.

The chemical procedures can be easily combined with lithography. In
Ref.~\citenum{Brongersma} the combination of lithography and
oxidization in air was successfully implemented for the fabrication
of a periodic Fe$_{2}$O$_{3}$ nanobeam-array for active
photocatalytic material creation supporting optical Mie resonances,
Fig.~\ref{fig:Fabr_2}(b). Chemical methods can also be combined with
standard commercial products to generate nanostructures for new
dielectric metamaterial designs, e.g. SiC whiskers available
commercially were washed in ethanol and studied~\cite{Schuller2007}.

Thus, chemical methods  provide large opportunities for researchers.
Whereas, inherent disadvantages (chemical waste, possible
contamination of fabricated nanomaterials, additional steps for
generation nanostructures, etc.) impose constraints on possible
application areas.

\subsection{Dewetting}

Dewetting of a thin film is another process that can be applied for
fabrication of high-index nanoparticles on a large scale. This
process implies agglomeration of nanoparticles during heating of a
thin film due to minimization of total energy of the thin film
surfaces, including a film-substrate
interface~\cite{ye2011templated,thompson2012solid}. The film
thickness has a direct influence on the dewetting process (the lower
the thickness, the higher the driving force for
dewetting)~\cite{thompson2012solid}, therefore dewetting can be
carried out at temperatures lower than the melting threshold of the
thin film material. Overall, the main controlling parameters in this method are the heating temperature and properties of the thin film
(thickness, presence of defects and initial patterns).

Dewetting has been applied for the fabrication of silicon
nanoparticles with different sizes after heating of thin
crystalline~\cite{AbbarchiACSNano2014} or
amorphous~\cite{AbbarchiNanoscale} silicon films. This method also enables
a controlled formation of complex assemblies of silicon
monocrystalline resonators, Fig.~\ref{fig:Fabr_2}(c)~\cite{AbbarchiACSNano2014}. It
should be noted, that in the thin film dewetting technique the
nanoparticle size and location control can be achieved only by using
additional lithographic methods, which is even more complicated
compared to the chemical deposition techniques.

In spite of this fact, dewetting offers great opportunities in terms
of productivity and can be applied to any material. In
Ref.~\citenum{Anderson} it was demonstrated, that the appropriate
choice of the substrate temperature during the growth process
provides an effective approach for the creation of needle-like Te
crystallites. In Ref.~\citenum{Lagally}, two-component SiGe
nanoparticles were obtained after thermal dewetting. In this work
thermal dewetting and agglomeration of the Ge layer deposited on a 9
nm thick Si (001) layer was investigated. It was demonstrated, that
Ge layer lowers the dewetting temperature and makes it possible the
creation of SiGe nanoparticles (see Fig.~\ref{fig:Fabr_2}(d)).

Thereby, dewetting is simple and highly productive method for the
fabrication of high-index nanoparticles. However, controllable
arrangement of nanoparticles  in certain places on the sample
surface is still a problem for this method.

\subsection{Laser-assisted methods}

Fast progress in nanotechnologies demands growing precision of
fabrication processes. It is not surprising, that laser-assisted
methods are  applied in different nanofabrication processes due to
their material selectivity, submicron resolution, high energy
density, etc. For example, laser-assisted methods were proven to be
effective  for the fabrication of nanoparticles with the diameters
less than 100~nm. It is worth noting that colloids of chemically
pure nanoparticles can be obtained by means of laser ablation in
liquids. The main advantages of the ablation approach are relatively
high productivity and lack of harmful chemical wastes.

The growing interest in the fabrication of high-index nanoparticles
stimulated application of laser-assisted methods for the generation
of nanoparticles with the sizes larger than 100~nm supporting Mie
resonances in the visible and near-IR regions. Studies in this field
have only recently begun and laser-assisted methods were applied
mostly to silicon.

\begin{figure}[!t] \centering
\includegraphics[width=1\columnwidth]{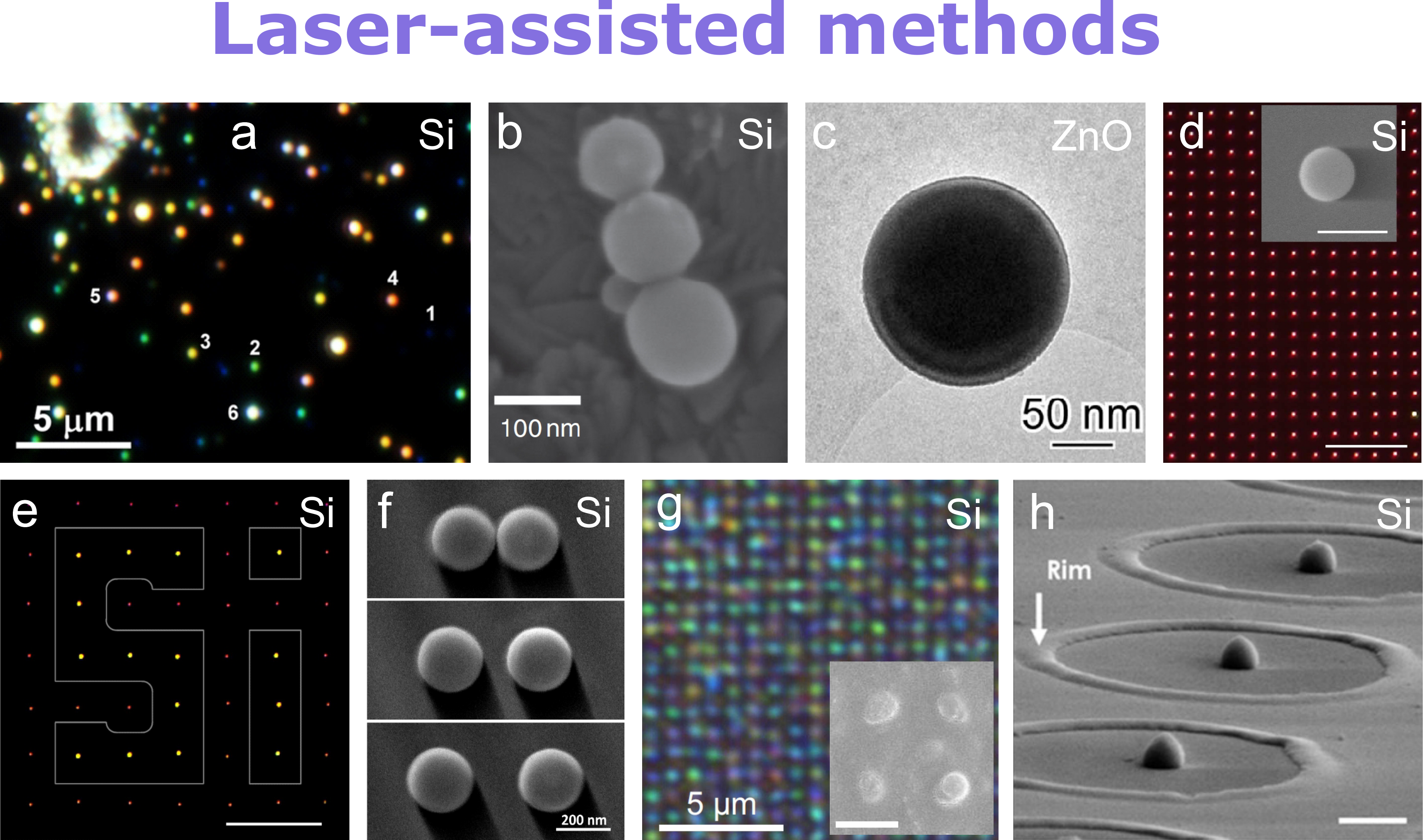}
\caption{(a) Dark-field optical image of silicon nanoparticles
obtained via femtosecond laser ablation of bulk
silicon, reprinted with permission from Ref.~\citenum{Kuznetsov2012}. (b) SEM image of a silicon nanosphere
trimer, reprinted with permission from Ref.~\citenum{Yang}. (c) TEM image of a ZnO submicrosphere, reprinted with permission from Ref.~\citenum{Okamoto}. (d-e) Demonstration of the laser printing method: (d) Array of amorphous Si nanoparticles fabricated by LIT method and visualized with dark-field microscopy (scale bar, 20 $\mu$m). The insert shows a SEM
image of a single Si nanoparticle in this array (scale bar, 200 nm). (e) Dark-field optical image of silicon nanoparticles obtained via femtosecond laser ablation of thin silicon film. In picture (e) red nanoparticles are amorphized, while yellow are annealed and crystalline. Reprinted with permission from Ref.~\citenum{Zywietz2014}.
(f) SEM-images of Si nanoparticle dimer structures on a glass substrate. Reprinted with permission from Ref.~\citenum{ZywietzACSPhoton2015}. (g) Optical image of an array of the silicon nanoparticles fabricated by direct laser writing, inset displays SEM image of the written nanoparticles covered by a 10~nm gold layer with the scale bar of 700 nm. Reprinted with permission from Ref.~\citenum{DmitrievNanoscale2015}.
(h) SEM pictures of silicon nanodome array produced by donut-shaped laser beam
pulses. Reprinted with permission from Ref.~\citenum{CostasNanotech2015}.}
\label{fig:Fabr_3}
\end{figure}

\subsubsection{Ablation of bulk materials}

First experiments towards fabrication of high-index nanoparticles
using laser-assisted methods were conducted by direct laser
ablation: an ultrashort laser pulse induced material fragmentation
into spherical nanoparticles and their deposition close to the focus
area, Fig.~\ref{fig:Fabr_3}(a)~\cite{Evlyukhin2012,Kuznetsov2012,
Fu2013,ZywietzAPA2014,Lewi2015}. Laser ablation in superfluid helium
was also successfully implemented for the fabrication of
single-crystalline sub- and micron-sized ZnO, CdSe, ZnSe, and
CeO$_2$ microspheres, Fig.~\ref{fig:Fabr_3}(b),(c)~\cite{Okamoto}.
These experiments proved the effectiveness of laser ablation for the
fabrication of high-index nanoparticles with optical response
(scattering efficiencies, Q-factors, etc.) in the visible and IR
spectral ranges. However, application of such processes in
nanophotonics is a problem due to inability to control the
fabricated nanoparticle sizes and their locations. To overcome these
problems, other laser-assisted methods were developed.

\subsubsection{Laser-induced transfer}

Laser-induced transfer (LIT) methods, demonstrated for the first
time in the 80s~\cite{bohandy1988metal}, has become a promising
approach for laser printing of nanoparticles from different
materials: metals and semiconductors. In this method, laser
radiation is focused on the interface between the printed material
and transparent donor substrate providing material transfer onto
another receiver substrate placed in a closed contact with the donor
sample. First experiments on the fabrication of silicon
nanoparticles with Mie-resonances in the visible range using laser
printing were performed in Ref.~\citenum{Zywietz2014}. With this
technique highly ordered arrays of nanoparticles can be
produced~\cite{Zywietz2014}, also dimers consisting of submicron
crystalline silicon nanoparticles with different interparticle
distances, ranging from ~5 nm to ~375 nm, have been demonstrated (
Fig.~\ref{fig:Fabr_3}(f)~\cite{ZywietzACSPhoton2015}). It should be
mentioned that laser annealing can be applied for postprocessing of
laser printed semiconductor nanoparticles to controllably change
their phase from initially amorphous state to crystalline. This
allows tailoring their optical properties, e.g., for silicon
nanoparticles, Fig.~\ref{fig:Fabr_3}(e)~\cite{Zywietz2014}.
Therefore, laser printing technique is very attractive for the
fabrication of high-index nanoparticles and their arrangement in 2D
arrays with a high precision.

\subsubsection{Laser-induced dewetting}

Laser radiation can be applied for patterning of thin films of
high-index materials by their controlled dewetting into nanoscale
structures. In Ref.~\citenum{DmitrievNanoscale2015}, crystalline
silicon nanoparticles were fabricated from amorphous silicon films
without applying initially crystalline materials or any additional
annealing steps (Fig.~\ref{fig:Fabr_3}(g)). Laser-based dewetting of
amorphous silicon thin films by donut-shaped laser beams was
demonstrated in Ref.~\citenum{CostasNanotech2015}, where
morphological modification of the film into a nanodome
(Fig.~\ref{fig:Fabr_3}(h)) during thermocapillary-induced dewetting
lead to phase transformation from amorphous to crystalline structure
in the laser focus area. Thus, laser-assisted dewetting combining
advantages of dewetting and direct laser patterning can be
considered as a promising method for large scale fabrication of
high-index nanoparticles.

To summarize, it should be noted that laser-assisted methods require
high quality laser radiation (high laser pulse stability, perfect
beam shape, and excellent focusing), including high precision and
reliability of the position systems. However, these constrains do
not lessen the basic advantages of the laser-assisted methods
(single step process, high repeatability, modification of
nanoparticles crystalline phase, space arrangement of the fabricated
nanoparticles), which all are important for the realization of
nanophotonic devices.

\section{Discussion and Outlook}
\begin{figure}
\includegraphics[width=.9\columnwidth]{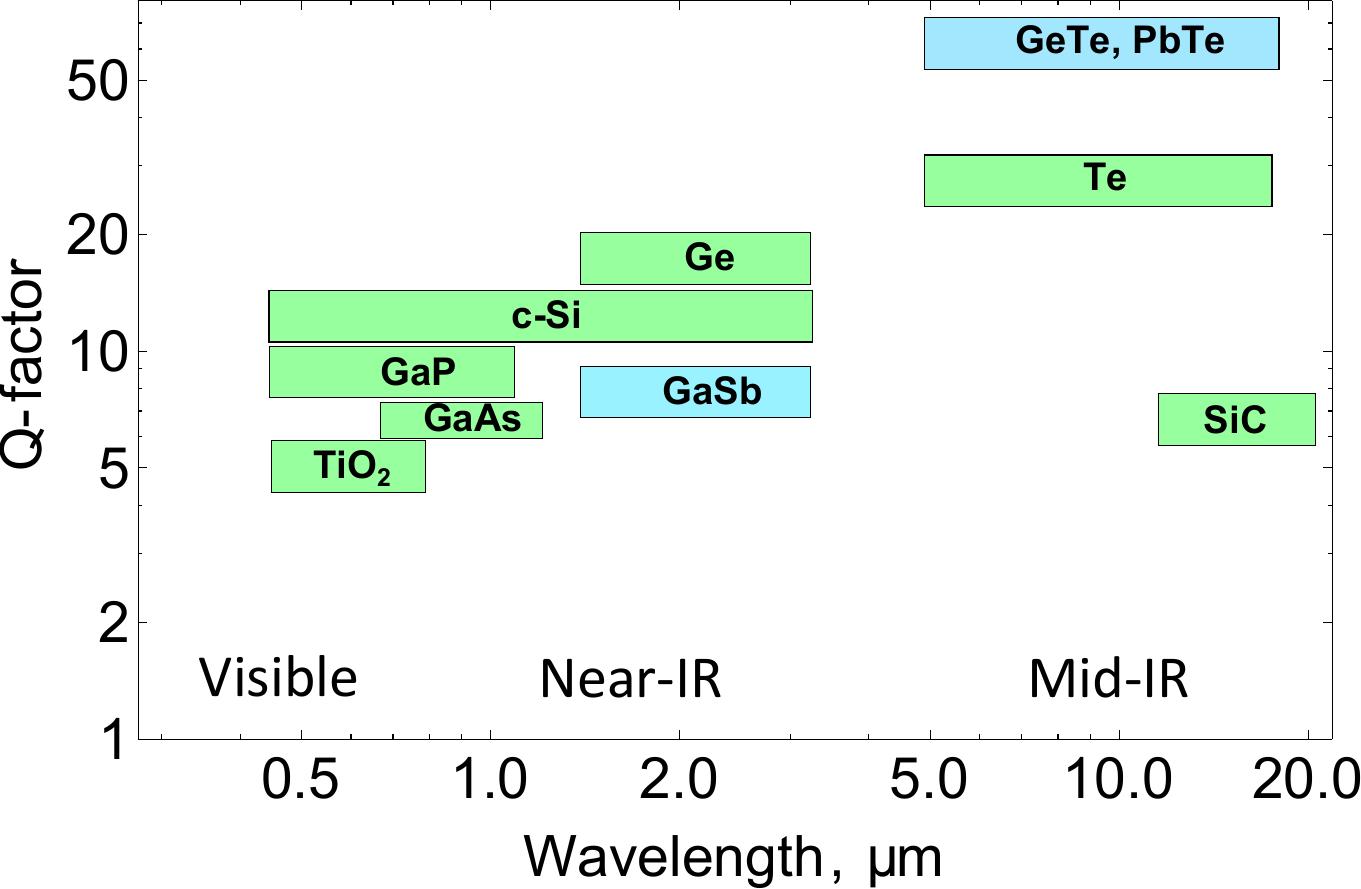}
\caption{High-refractive index materials for all-dielectric
nanophotonics: $Q$-factors of spherical resonators made from these
materials. Materials which can be nanostructured by the available
fabrication techniques are marked by green, non-structurable
materials are marked by blue color.} \label{fig4}
\end{figure}

An overview of the available high-index materials and
nanofabrication methods for their potential applications in
all-dielectric nanophotonics has been given. Figure~\ref{fig4}
briefly summarizes our  analysis of high-refractive index materials.
In this plot, materials for which nanoparticle fabrication methods
have been demonstrated are highlighted by green color, while
materials, for which nanofabrication has not been demonstrated so
far, are highlighted by blue color. Optical performance of each
material in terms of the resonance quality factor is determined by
the trade-off between the refractive index and related optical
absorption.

In the visible region, silicon nanoantennas provide highest field
enhancement, while GaP may be preferable for the realization of
metasurfaces due to smaller absorption. In the near-IR region,
germanium is the best high-index material, while the narrow band-gap
semiconductors, such as GeTe and PbTe, are more promising for the
mid-IR range.

In terms of nanofabrication, currently developed approaches allow
production of different types of all-dielectric functional
structures with the desired optical properties. The development of
hybrid nanostructures combining plasmonic and Mie resonances will be
interesting for different applications in nanophotonics, medicine,
ecology, etc.
% result

Further research in this field can be devoted to the search for
other materials demonstrating superior behavior in the visible and
near-IR regions. Consideration of indirect band gap semiconductors,
where optical absorption is suppressed due to the mismatch between
photon and electron wavevectors, is promising. We think that the
ideal high-index material could be found among narrow-gap indirect
semiconductors.

It is also possible to use gapless direct semiconductors or
semimetals, including 3D Dirac semimetals~\cite{Cava, Dirac,
Topological}, where absorption is suppressed due to optical Pauli
blocking. However, for applications in the visible range, one needs
materials with very high Fermi level ($E_{\rm f}>1$~eV). Optical
constants are not determined solely by the electronic band
structure. They can be  modified by phonon-polariton resonances.
These resonances can give a higher refractive index simultaneously
bringing losses. Large density of phonon modes at room temperature
prevents currently known gapless semiconductors and 3D Dirac
semimetals from becoming candidates for lossless high-index
materials. On the other hand, specific fabrication techniques should
be developed to explore potential of narrow band gap semiconductors,
such as PbTe and GeTe, as materials for sub-wavelength mid-IR
resonators.

We hope that the provided analysis will serve for better
understanding of design rules of high-index nanoresonators and will
facilitate the development of highly-efficient optical devices.

\begin{acknowledgments}
Authors acknowledge fruitful discussion with Philippe Tassin, Boris Luk'yanchuk, Andrea Alu, Pavel Belov, and Yuri Kivshar.
%D.G.B. acknowledges support from the Russian Foundation for Basic Research (project No 16-32-00444).
\end{acknowledgments}

$^\dag$ These authors contributed equally to this manuscript.

\bibliography{high_index,fabric}

\begin{thebibliography}{113}%
\makeatletter
\providecommand \@ifxundefined [1]{%
 \@ifx{#1\undefined}
}%
\providecommand \@ifnum [1]{%
 \ifnum #1\expandafter \@firstoftwo
 \else \expandafter \@secondoftwo
 \fi
}%
\providecommand \@ifx [1]{%
 \ifx #1\expandafter \@firstoftwo
 \else \expandafter \@secondoftwo
 \fi
}%
\providecommand \natexlab [1]{#1}%
\providecommand \enquote  [1]{``#1''}%
\providecommand \bibnamefont  [1]{#1}%
\providecommand \bibfnamefont [1]{#1}%
\providecommand \citenamefont [1]{#1}%
\providecommand \href@noop [0]{\@secondoftwo}%
\providecommand \href [0]{\begingroup \@sanitize@url \@href}%
\providecommand \@href[1]{\@@startlink{#1}\@@href}%
\providecommand \@@href[1]{\endgroup#1\@@endlink}%
\providecommand \@sanitize@url [0]{\catcode `\\12\catcode `\$12\catcode
  `\&12\catcode `\#12\catcode `\^12\catcode `\_12\catcode `\%12\relax}%
\providecommand \@@startlink[1]{}%
\providecommand \@@endlink[0]{}%
\providecommand \url  [0]{\begingroup\@sanitize@url \@url }%
\providecommand \@url [1]{\endgroup\@href {#1}{\urlprefix }}%
\providecommand \urlprefix  [0]{URL }%
\providecommand \Eprint [0]{\href }%
\providecommand \doibase [0]{http://dx.doi.org/}%
\providecommand \selectlanguage [0]{\@gobble}%
\providecommand \bibinfo  [0]{\@secondoftwo}%
\providecommand \bibfield  [0]{\@secondoftwo}%
\providecommand \translation [1]{[#1]}%
\providecommand \BibitemOpen [0]{}%
\providecommand \bibitemStop [0]{}%
\providecommand \bibitemNoStop [0]{.\EOS\space}%
\providecommand \EOS [0]{\spacefactor3000\relax}%
\providecommand \BibitemShut  [1]{\csname bibitem#1\endcsname}%
\let\auto@bib@innerbib\@empty
%</preamble>
\bibitem [{\citenamefont {Pelton}\ \emph {et~al.}(2008)\citenamefont {Pelton},
  \citenamefont {Aizpurua},\ and\ \citenamefont {Bryant}}]{Pelton2008}%
  \BibitemOpen
  \bibfield  {author} {\bibinfo {author} {\bibfnamefont {M.}~\bibnamefont
  {Pelton}}, \bibinfo {author} {\bibfnamefont {J.}~\bibnamefont {Aizpurua}}, \
  and\ \bibinfo {author} {\bibfnamefont {G.}~\bibnamefont {Bryant}},\ }\href
  {\doibase 10.1002/lpor.200810003} {\bibfield  {journal} {\bibinfo  {journal}
  {Laser Photon. Rev.}\ }\textbf {\bibinfo {volume} {2}},\ \bibinfo {pages}
  {136} (\bibinfo {year} {2008})}\BibitemShut {NoStop}%
\bibitem [{\citenamefont {Schuller}\ \emph {et~al.}(2010)\citenamefont
  {Schuller}, \citenamefont {Barnard}, \citenamefont {Cai}, \citenamefont
  {Jun}, \citenamefont {White},\ and\ \citenamefont
  {Brongersma}}]{Schuller2010}%
  \BibitemOpen
  \bibfield  {author} {\bibinfo {author} {\bibfnamefont {J.~A.}\ \bibnamefont
  {Schuller}}, \bibinfo {author} {\bibfnamefont {E.~S.}\ \bibnamefont
  {Barnard}}, \bibinfo {author} {\bibfnamefont {W.}~\bibnamefont {Cai}},
  \bibinfo {author} {\bibfnamefont {Y.~C.}\ \bibnamefont {Jun}}, \bibinfo
  {author} {\bibfnamefont {J.~S.}\ \bibnamefont {White}}, \ and\ \bibinfo
  {author} {\bibfnamefont {M.~L.}\ \bibnamefont {Brongersma}},\ }\href@noop {}
  {\bibfield  {journal} {\bibinfo  {journal} {Nat. Mater.}\ }\textbf {\bibinfo
  {volume} {9}},\ \bibinfo {pages} {193} (\bibinfo {year} {2010})}\BibitemShut
  {NoStop}%
\bibitem [{\citenamefont {Giannini}\ \emph {et~al.}(2011)\citenamefont
  {Giannini}, \citenamefont {Fernandez-Dominguez}, \citenamefont {Heck},\ and\
  \citenamefont {Maier}}]{Giannini2011}%
  \BibitemOpen
  \bibfield  {author} {\bibinfo {author} {\bibfnamefont {V.}~\bibnamefont
  {Giannini}}, \bibinfo {author} {\bibfnamefont {A.~I.}\ \bibnamefont
  {Fernandez-Dominguez}}, \bibinfo {author} {\bibfnamefont {S.~C.}\
  \bibnamefont {Heck}}, \ and\ \bibinfo {author} {\bibfnamefont {S.~A.}\
  \bibnamefont {Maier}},\ }\href {\doibase 10.1021/cr1002672} {\bibfield
  {journal} {\bibinfo  {journal} {Chemical Reviews}\ }\textbf {\bibinfo
  {volume} {111}},\ \bibinfo {pages} {3888} (\bibinfo {year}
  {2011})}\BibitemShut {NoStop}%
\bibitem [{\citenamefont {Fan}\ \emph {et~al.}(2014)\citenamefont {Fan},
  \citenamefont {Zheng},\ and\ \citenamefont {Singh}}]{Fan2014}%
  \BibitemOpen
  \bibfield  {author} {\bibinfo {author} {\bibfnamefont {X.}~\bibnamefont
  {Fan}}, \bibinfo {author} {\bibfnamefont {W.}~\bibnamefont {Zheng}}, \ and\
  \bibinfo {author} {\bibfnamefont {D.~J.}\ \bibnamefont {Singh}},\ }\href
  {\doibase 10.1038/lsa.2014.60} {\bibfield  {journal} {\bibinfo  {journal}
  {Light Sci. Appl.}\ }\textbf {\bibinfo {volume} {3}},\ \bibinfo {pages}
  {e179} (\bibinfo {year} {2014})}\BibitemShut {NoStop}%
\bibitem [{\citenamefont {West}\ \emph {et~al.}(2010)\citenamefont {West},
  \citenamefont {Ishii}, \citenamefont {Naik}, \citenamefont {Emani},
  \citenamefont {Shalaev},\ and\ \citenamefont {Boltasseva}}]{West2010}%
  \BibitemOpen
  \bibfield  {author} {\bibinfo {author} {\bibfnamefont {P.~R.}\ \bibnamefont
  {West}}, \bibinfo {author} {\bibfnamefont {S.}~\bibnamefont {Ishii}},
  \bibinfo {author} {\bibfnamefont {G.~V.}\ \bibnamefont {Naik}}, \bibinfo
  {author} {\bibfnamefont {N.~K.}\ \bibnamefont {Emani}}, \bibinfo {author}
  {\bibfnamefont {V.~M.}\ \bibnamefont {Shalaev}}, \ and\ \bibinfo {author}
  {\bibfnamefont {A.}~\bibnamefont {Boltasseva}},\ }\href {\doibase
  10.1002/lpor.200900055} {\bibfield  {journal} {\bibinfo  {journal} {Laser
  Photon. Rev.}\ }\textbf {\bibinfo {volume} {4}},\ \bibinfo {pages} {795}
  (\bibinfo {year} {2010})}\BibitemShut {NoStop}%
\bibitem [{\citenamefont {Khurgin}\ and\ \citenamefont
  {Sun}(2012)}]{Khurgin2012}%
  \BibitemOpen
  \bibfield  {author} {\bibinfo {author} {\bibfnamefont {J.~B.}\ \bibnamefont
  {Khurgin}}\ and\ \bibinfo {author} {\bibfnamefont {G.}~\bibnamefont {Sun}},\
  }\href@noop {} {\bibfield  {journal} {\bibinfo  {journal} {App. Phys. Lett.}\
  }\textbf {\bibinfo {volume} {100}},\ \bibinfo {pages} {11105} (\bibinfo
  {year} {2012})}\BibitemShut {NoStop}%
\bibitem [{\citenamefont {Khurgin}(2015)}]{Khurgin2015}%
  \BibitemOpen
  \bibfield  {author} {\bibinfo {author} {\bibfnamefont {J.~B.}\ \bibnamefont
  {Khurgin}},\ }\href {\doibase 10.1038/nnano.2014.310} {\bibfield  {journal}
  {\bibinfo  {journal} {Nat. Nanotechnol.}\ }\textbf {\bibinfo {volume} {10}},\
  \bibinfo {pages} {2} (\bibinfo {year} {2015})}\BibitemShut {NoStop}%
\bibitem [{\citenamefont {Gather}\ \emph {et~al.}(2010)\citenamefont {Gather},
  \citenamefont {Meerholz}, \citenamefont {Danz},\ and\ \citenamefont
  {Leosson}}]{Gather2010}%
  \BibitemOpen
  \bibfield  {author} {\bibinfo {author} {\bibfnamefont {M.}~\bibnamefont
  {Gather}}, \bibinfo {author} {\bibfnamefont {K.}~\bibnamefont {Meerholz}},
  \bibinfo {author} {\bibfnamefont {N.}~\bibnamefont {Danz}}, \ and\ \bibinfo
  {author} {\bibfnamefont {K.}~\bibnamefont {Leosson}},\ }\href {\doibase
  10.1038/NPHOTON.2010.121} {\bibfield  {journal} {\bibinfo  {journal} {Nature
  Photon.}\ }\textbf {\bibinfo {volume} {4}},\ \bibinfo {pages} {457} (\bibinfo
  {year} {2010})}\BibitemShut {NoStop}%
\bibitem [{\citenamefont {Wuestner}\ \emph {et~al.}(2010)\citenamefont
  {Wuestner}, \citenamefont {Pusch}, \citenamefont {Tsakmakidis}, \citenamefont
  {Hamm},\ and\ \citenamefont {Hess}}]{Wuestner2010}%
  \BibitemOpen
  \bibfield  {author} {\bibinfo {author} {\bibfnamefont {S.}~\bibnamefont
  {Wuestner}}, \bibinfo {author} {\bibfnamefont {A.}~\bibnamefont {Pusch}},
  \bibinfo {author} {\bibfnamefont {K.~L.}\ \bibnamefont {Tsakmakidis}},
  \bibinfo {author} {\bibfnamefont {J.~M.}\ \bibnamefont {Hamm}}, \ and\
  \bibinfo {author} {\bibfnamefont {O.}~\bibnamefont {Hess}},\ }\href {\doibase
  10.1103/PhysRevLett.105.127401} {\bibfield  {journal} {\bibinfo  {journal}
  {Phys. Rev. Lett.}\ }\textbf {\bibinfo {volume} {105}},\ \bibinfo {pages}
  {127401} (\bibinfo {year} {2010})}\BibitemShut {NoStop}%
\bibitem [{\citenamefont {Stockman}(2011)}]{Stockman2011}%
  \BibitemOpen
  \bibfield  {author} {\bibinfo {author} {\bibfnamefont {M.~I.}\ \bibnamefont
  {Stockman}},\ }\href {http://www.ncbi.nlm.nih.gov/pubmed/21568593} {\bibfield
   {journal} {\bibinfo  {journal} {Phys. Rev. Lett.}\ }\textbf {\bibinfo
  {volume} {106}},\ \bibinfo {pages} {156802} (\bibinfo {year}
  {2011})}\BibitemShut {NoStop}%
\bibitem [{\citenamefont {Pusch}\ \emph {et~al.}(2012)\citenamefont {Pusch},
  \citenamefont {Wuestner}, \citenamefont {Hamm}, \citenamefont {Tsakmakidis},
  \citenamefont {Hess}, \citenamefont {Maxwell-bloch}, \citenamefont {Pusch},
  \citenamefont {Wuestner}, \citenamefont {Hamm}, \citenamefont {Tsakmakidis},\
  and\ \citenamefont {Hess}}]{Pusch2012}%
  \BibitemOpen
  \bibfield  {author} {\bibinfo {author} {\bibfnamefont {A.}~\bibnamefont
  {Pusch}}, \bibinfo {author} {\bibfnamefont {S.}~\bibnamefont {Wuestner}},
  \bibinfo {author} {\bibfnamefont {J.~M.}\ \bibnamefont {Hamm}}, \bibinfo
  {author} {\bibfnamefont {K.~L.}\ \bibnamefont {Tsakmakidis}}, \bibinfo
  {author} {\bibfnamefont {O.}~\bibnamefont {Hess}}, \bibinfo {author}
  {\bibfnamefont {M.~A.}\ \bibnamefont {Maxwell-bloch}}, \bibinfo {author}
  {\bibfnamefont {A.}~\bibnamefont {Pusch}}, \bibinfo {author} {\bibfnamefont
  {S.}~\bibnamefont {Wuestner}}, \bibinfo {author} {\bibfnamefont {J.~M.}\
  \bibnamefont {Hamm}}, \bibinfo {author} {\bibfnamefont {K.~L.}\ \bibnamefont
  {Tsakmakidis}}, \ and\ \bibinfo {author} {\bibfnamefont {O.}~\bibnamefont
  {Hess}},\ }\href {\doibase 10.1021/nn204692x} {\bibfield  {journal} {\bibinfo
   {journal} {ACS Nano}\ }\textbf {\bibinfo {volume} {6}},\ \bibinfo {pages}
  {2420} (\bibinfo {year} {2012})}\BibitemShut {NoStop}%
\bibitem [{\citenamefont {Lisyansky}\ \emph {et~al.}(2011)\citenamefont
  {Lisyansky}, \citenamefont {Nechepurenko}, \citenamefont {Dorofeenko},
  \citenamefont {Vinogradov},\ and\ \citenamefont {Pukhov}}]{Lisyansky2011}%
  \BibitemOpen
  \bibfield  {author} {\bibinfo {author} {\bibfnamefont {A.~A.}\ \bibnamefont
  {Lisyansky}}, \bibinfo {author} {\bibfnamefont {I.~A.}\ \bibnamefont
  {Nechepurenko}}, \bibinfo {author} {\bibfnamefont {A.~V.}\ \bibnamefont
  {Dorofeenko}}, \bibinfo {author} {\bibfnamefont {A.~P.}\ \bibnamefont
  {Vinogradov}}, \ and\ \bibinfo {author} {\bibfnamefont {A.~A.}\ \bibnamefont
  {Pukhov}},\ }\href {\doibase 10.1103/PhysRevB.84.153409} {\bibfield
  {journal} {\bibinfo  {journal} {Phys. Rev. B}\ }\textbf {\bibinfo {volume}
  {84}},\ \bibinfo {pages} {153409} (\bibinfo {year} {2011})}\BibitemShut
  {NoStop}%
\bibitem [{\citenamefont {Andrianov}\ \emph {et~al.}(2013)\citenamefont
  {Andrianov}, \citenamefont {Baranov}, \citenamefont {Pukhov}, \citenamefont
  {Dorofeenko}, \citenamefont {Vinogradov},\ and\ \citenamefont
  {Lisyansky}}]{Andrianov2013}%
  \BibitemOpen
  \bibfield  {author} {\bibinfo {author} {\bibfnamefont {E.~S.}\ \bibnamefont
  {Andrianov}}, \bibinfo {author} {\bibfnamefont {D.~G.}\ \bibnamefont
  {Baranov}}, \bibinfo {author} {\bibfnamefont {A.~A.}\ \bibnamefont {Pukhov}},
  \bibinfo {author} {\bibfnamefont {A.~V.}\ \bibnamefont {Dorofeenko}},
  \bibinfo {author} {\bibfnamefont {A.~P.}\ \bibnamefont {Vinogradov}}, \ and\
  \bibinfo {author} {\bibfnamefont {A.~A.}\ \bibnamefont {Lisyansky}},\ }\href
  {http://www.ncbi.nlm.nih.gov/pubmed/23736600} {\bibfield  {journal} {\bibinfo
   {journal} {Opt. Express}\ }\textbf {\bibinfo {volume} {21}},\ \bibinfo
  {pages} {13467} (\bibinfo {year} {2013})}\BibitemShut {NoStop}%
\bibitem [{\citenamefont {Khurgin}\ and\ \citenamefont
  {Sun}(2010)}]{Khurgin2010}%
  \BibitemOpen
  \bibfield  {author} {\bibinfo {author} {\bibfnamefont {J.~B.}\ \bibnamefont
  {Khurgin}}\ and\ \bibinfo {author} {\bibfnamefont {G.}~\bibnamefont {Sun}},\
  }\href@noop {} {\bibfield  {journal} {\bibinfo  {journal} {Appl. Phys.
  Lett.}\ }\textbf {\bibinfo {volume} {96}},\ \bibinfo {pages} {181102}
  (\bibinfo {year} {2010})}\BibitemShut {NoStop}%
\bibitem [{\citenamefont {Naik}\ \emph {et~al.}(2013)\citenamefont {Naik},
  \citenamefont {Shalaev},\ and\ \citenamefont {Boltasseva}}]{Naik2013}%
  \BibitemOpen
  \bibfield  {author} {\bibinfo {author} {\bibfnamefont {G.~V.}\ \bibnamefont
  {Naik}}, \bibinfo {author} {\bibfnamefont {V.~M.}\ \bibnamefont {Shalaev}}, \
  and\ \bibinfo {author} {\bibfnamefont {A.}~\bibnamefont {Boltasseva}},\
  }\href@noop {} {\bibfield  {journal} {\bibinfo  {journal} {Adv. Mater.}\
  }\textbf {\bibinfo {volume} {25}},\ \bibinfo {pages} {3264} (\bibinfo {year}
  {2013})}\BibitemShut {NoStop}%
\bibitem [{\citenamefont {Feng}\ \emph {et~al.}(2015)\citenamefont {Feng},
  \citenamefont {Streyer}, \citenamefont {Zhong}, \citenamefont {Hoffman},\
  and\ \citenamefont {Wasserman}}]{Feng2015}%
  \BibitemOpen
  \bibfield  {author} {\bibinfo {author} {\bibfnamefont {K.}~\bibnamefont
  {Feng}}, \bibinfo {author} {\bibfnamefont {W.}~\bibnamefont {Streyer}},
  \bibinfo {author} {\bibfnamefont {Y.}~\bibnamefont {Zhong}}, \bibinfo
  {author} {\bibfnamefont {A.}~\bibnamefont {Hoffman}}, \ and\ \bibinfo
  {author} {\bibfnamefont {D.}~\bibnamefont {Wasserman}},\ }\href {\doibase
  10.1364/OE.23.0A1418} {\bibfield  {journal} {\bibinfo  {journal} {Opt.
  Express}\ }\textbf {\bibinfo {volume} {23}},\ \bibinfo {pages} {A1418}
  (\bibinfo {year} {2015})}\BibitemShut {NoStop}%
\bibitem [{\citenamefont {Mie}(1908)}]{Mie}%
  \BibitemOpen
  \bibfield  {author} {\bibinfo {author} {\bibfnamefont {G.}~\bibnamefont
  {Mie}},\ }\href {\doibase 10.1002/andp.19083300302} {\bibfield  {journal}
  {\bibinfo  {journal} {Ann. Phys. (Berlin)}\ }\textbf {\bibinfo {volume}
  {330}},\ \bibinfo {pages} {377} (\bibinfo {year} {1908})}\BibitemShut
  {NoStop}%
\bibitem [{\citenamefont {Fenollosa}\ \emph {et~al.}(2008)\citenamefont
  {Fenollosa}, \citenamefont {Meseguer},\ and\ \citenamefont
  {Tymczenko}}]{Fenollosa}%
  \BibitemOpen
  \bibfield  {author} {\bibinfo {author} {\bibfnamefont {R.}~\bibnamefont
  {Fenollosa}}, \bibinfo {author} {\bibfnamefont {F.}~\bibnamefont {Meseguer}},
  \ and\ \bibinfo {author} {\bibfnamefont {M.}~\bibnamefont {Tymczenko}},\
  }\href@noop {} {\bibfield  {journal} {\bibinfo  {journal} {Adv. Mater.}\
  }\textbf {\bibinfo {volume} {20}},\ \bibinfo {pages} {95} (\bibinfo {year}
  {2008})}\BibitemShut {NoStop}%
\bibitem [{\citenamefont {Xifre-Perez}\ \emph {et~al.}(2011)\citenamefont
  {Xifre-Perez}, \citenamefont {Fenollosa},\ and\ \citenamefont
  {Meseguer}}]{Xifre}%
  \BibitemOpen
  \bibfield  {author} {\bibinfo {author} {\bibfnamefont {E.}~\bibnamefont
  {Xifre-Perez}}, \bibinfo {author} {\bibfnamefont {R.}~\bibnamefont
  {Fenollosa}}, \ and\ \bibinfo {author} {\bibfnamefont {F.}~\bibnamefont
  {Meseguer}},\ }\href@noop {} {\bibfield  {journal} {\bibinfo  {journal} {Opt.
  Express}\ }\textbf {\bibinfo {volume} {19}},\ \bibinfo {pages} {3455}
  (\bibinfo {year} {2011})}\BibitemShut {NoStop}%
\bibitem [{\citenamefont {Evlyukhin}\ \emph {et~al.}(2012)\citenamefont
  {Evlyukhin}, \citenamefont {Novikov}, \citenamefont {Zywietz}, \citenamefont
  {Eriksen}, \citenamefont {Reinhardt}, \citenamefont {Bozhevolnyi},\ and\
  \citenamefont {Chichkov}}]{Evlyukhin2012}%
  \BibitemOpen
  \bibfield  {author} {\bibinfo {author} {\bibfnamefont {A.~B.}\ \bibnamefont
  {Evlyukhin}}, \bibinfo {author} {\bibfnamefont {S.~M.}\ \bibnamefont
  {Novikov}}, \bibinfo {author} {\bibfnamefont {U.}~\bibnamefont {Zywietz}},
  \bibinfo {author} {\bibfnamefont {R.~L.}\ \bibnamefont {Eriksen}}, \bibinfo
  {author} {\bibfnamefont {C.}~\bibnamefont {Reinhardt}}, \bibinfo {author}
  {\bibfnamefont {S.~I.}\ \bibnamefont {Bozhevolnyi}}, \ and\ \bibinfo {author}
  {\bibfnamefont {B.~N.}\ \bibnamefont {Chichkov}},\ }\href {\doibase
  10.1021/nl301594s} {\bibfield  {journal} {\bibinfo  {journal} {Nano Lett.}\
  }\textbf {\bibinfo {volume} {12}},\ \bibinfo {pages} {3749} (\bibinfo {year}
  {2012})}\BibitemShut {NoStop}%
\bibitem [{\citenamefont {Kuznetsov}\ \emph {et~al.}(2012)\citenamefont
  {Kuznetsov}, \citenamefont {Miroshnichenko}, \citenamefont {Fu},
  \citenamefont {Zhang},\ and\ \citenamefont {Luk'yanchuk}}]{Kuznetsov2012}%
  \BibitemOpen
  \bibfield  {author} {\bibinfo {author} {\bibfnamefont {A.~I.}\ \bibnamefont
  {Kuznetsov}}, \bibinfo {author} {\bibfnamefont {A.~E.}\ \bibnamefont
  {Miroshnichenko}}, \bibinfo {author} {\bibfnamefont {Y.~H.}\ \bibnamefont
  {Fu}}, \bibinfo {author} {\bibfnamefont {J.}~\bibnamefont {Zhang}}, \ and\
  \bibinfo {author} {\bibfnamefont {B.}~\bibnamefont {Luk'yanchuk}},\ }\href
  {\doibase 10.1038/srep00492} {\bibfield  {journal} {\bibinfo  {journal} {Sci.
  Rep.}\ }\textbf {\bibinfo {volume} {2}},\ \bibinfo {pages} {492} (\bibinfo
  {year} {2012})}\BibitemShut {NoStop}%
\bibitem [{\citenamefont {Zywietz}\ \emph
  {et~al.}(2014{\natexlab{a}})\citenamefont {Zywietz}, \citenamefont
  {Evlyukhin}, \citenamefont {Reinhardt},\ and\ \citenamefont
  {Chichkov}}]{Zywietz2014}%
  \BibitemOpen
  \bibfield  {author} {\bibinfo {author} {\bibfnamefont {U.}~\bibnamefont
  {Zywietz}}, \bibinfo {author} {\bibfnamefont {A.~B.}\ \bibnamefont
  {Evlyukhin}}, \bibinfo {author} {\bibfnamefont {C.}~\bibnamefont
  {Reinhardt}}, \ and\ \bibinfo {author} {\bibfnamefont {B.~N.}\ \bibnamefont
  {Chichkov}},\ }\href {\doibase 10.1038/ncomms4402} {\bibfield  {journal}
  {\bibinfo  {journal} {Nat. Commun.}\ }\textbf {\bibinfo {volume} {5}},\
  \bibinfo {pages} {3402} (\bibinfo {year} {2014}{\natexlab{a}})}\BibitemShut
  {NoStop}%
\bibitem [{\citenamefont {Jahani}\ and\ \citenamefont
  {Jacob}(2016)}]{Jahani2016}%
  \BibitemOpen
  \bibfield  {author} {\bibinfo {author} {\bibfnamefont {S.}~\bibnamefont
  {Jahani}}\ and\ \bibinfo {author} {\bibfnamefont {Z.}~\bibnamefont {Jacob}},\
  }\href {\doibase 10.1038/nnano.2015.304} {\bibfield  {journal} {\bibinfo
  {journal} {Nat. Nanotechnol.}\ }\textbf {\bibinfo {volume} {11}},\ \bibinfo
  {pages} {23} (\bibinfo {year} {2016})}\BibitemShut {NoStop}%
\bibitem [{\citenamefont {Kuznetsov}\ \emph {et~al.}(2016)\citenamefont
  {Kuznetsov}, \citenamefont {Miroshnichenko}, \citenamefont {Brongersma},
  \citenamefont {Kivshar},\ and\ \citenamefont {Lukyanchuk}}]{kuznetsov2016}%
  \BibitemOpen
  \bibfield  {author} {\bibinfo {author} {\bibfnamefont {A.~I.}\ \bibnamefont
  {Kuznetsov}}, \bibinfo {author} {\bibfnamefont {A.~E.}\ \bibnamefont
  {Miroshnichenko}}, \bibinfo {author} {\bibfnamefont {M.~L.}\ \bibnamefont
  {Brongersma}}, \bibinfo {author} {\bibfnamefont {Y.~S.}\ \bibnamefont
  {Kivshar}}, \ and\ \bibinfo {author} {\bibfnamefont {B.}~\bibnamefont
  {Lukyanchuk}},\ }\href@noop {} {\bibfield  {journal} {\bibinfo  {journal}
  {Science}\ }\textbf {\bibinfo {volume} {354}},\ \bibinfo {pages} {aag2472}
  (\bibinfo {year} {2016})}\BibitemShut {NoStop}%
\bibitem [{\citenamefont {Wheeler}\ \emph {et~al.}(2005)\citenamefont
  {Wheeler}, \citenamefont {Aitchison},\ and\ \citenamefont
  {Mojahedi}}]{Wheeler2005}%
  \BibitemOpen
  \bibfield  {author} {\bibinfo {author} {\bibfnamefont {M.~S.}\ \bibnamefont
  {Wheeler}}, \bibinfo {author} {\bibfnamefont {J.~S.}\ \bibnamefont
  {Aitchison}}, \ and\ \bibinfo {author} {\bibfnamefont {M.}~\bibnamefont
  {Mojahedi}},\ }\href {\doibase 10.1103/PhysRevB.72.193103} {\bibfield
  {journal} {\bibinfo  {journal} {Phys. Rev. B}\ }\textbf {\bibinfo {volume}
  {72}},\ \bibinfo {pages} {193103} (\bibinfo {year} {2005})}\BibitemShut
  {NoStop}%
\bibitem [{\citenamefont {Popa}\ and\ \citenamefont {Cummer}(2008)}]{Popa2008}%
  \BibitemOpen
  \bibfield  {author} {\bibinfo {author} {\bibfnamefont {B.~I.}\ \bibnamefont
  {Popa}}\ and\ \bibinfo {author} {\bibfnamefont {S.~A.}\ \bibnamefont
  {Cummer}},\ }\href {\doibase 10.1103/PhysRevLett.100.207401} {\bibfield
  {journal} {\bibinfo  {journal} {Phys. Rev. Lett.}\ }\textbf {\bibinfo
  {volume} {100}},\ \bibinfo {pages} {1} (\bibinfo {year} {2008})}\BibitemShut
  {NoStop}%
\bibitem [{\citenamefont {Schuller}\ \emph {et~al.}(2007)\citenamefont
  {Schuller}, \citenamefont {Zia}, \citenamefont {Taubner},\ and\ \citenamefont
  {Brongersma}}]{Schuller2007}%
  \BibitemOpen
  \bibfield  {author} {\bibinfo {author} {\bibfnamefont {J.~A.}\ \bibnamefont
  {Schuller}}, \bibinfo {author} {\bibfnamefont {R.}~\bibnamefont {Zia}},
  \bibinfo {author} {\bibfnamefont {T.}~\bibnamefont {Taubner}}, \ and\
  \bibinfo {author} {\bibfnamefont {M.~L.}\ \bibnamefont {Brongersma}},\ }\href
  {\doibase 10.1103/PhysRevLett.99.107401} {\bibfield  {journal} {\bibinfo
  {journal} {Phys. Rev. Lett.}\ }\textbf {\bibinfo {volume} {99}},\ \bibinfo
  {pages} {1} (\bibinfo {year} {2007})}\BibitemShut {NoStop}%
\bibitem [{\citenamefont {Ginn}\ \emph {et~al.}(2012)\citenamefont {Ginn},
  \citenamefont {Brener}, \citenamefont {Peters}, \citenamefont {Wendt},
  \citenamefont {Stevens}, \citenamefont {Hines}, \citenamefont {Basilio},
  \citenamefont {Warne}, \citenamefont {Ihlefeld}, \citenamefont {Clem},\ and\
  \citenamefont {Sinclair}}]{Ginn2012}%
  \BibitemOpen
  \bibfield  {author} {\bibinfo {author} {\bibfnamefont {J.~C.}\ \bibnamefont
  {Ginn}}, \bibinfo {author} {\bibfnamefont {I.}~\bibnamefont {Brener}},
  \bibinfo {author} {\bibfnamefont {D.~W.}\ \bibnamefont {Peters}}, \bibinfo
  {author} {\bibfnamefont {J.~R.}\ \bibnamefont {Wendt}}, \bibinfo {author}
  {\bibfnamefont {J.~O.}\ \bibnamefont {Stevens}}, \bibinfo {author}
  {\bibfnamefont {P.~F.}\ \bibnamefont {Hines}}, \bibinfo {author}
  {\bibfnamefont {L.~I.}\ \bibnamefont {Basilio}}, \bibinfo {author}
  {\bibfnamefont {L.~K.}\ \bibnamefont {Warne}}, \bibinfo {author}
  {\bibfnamefont {J.~F.}\ \bibnamefont {Ihlefeld}}, \bibinfo {author}
  {\bibfnamefont {P.~G.}\ \bibnamefont {Clem}}, \ and\ \bibinfo {author}
  {\bibfnamefont {M.~B.}\ \bibnamefont {Sinclair}},\ }\href {\doibase
  10.1103/PhysRevLett.108.097402} {\bibfield  {journal} {\bibinfo  {journal}
  {Phys. Rev. Lett.}\ }\textbf {\bibinfo {volume} {108}},\ \bibinfo {pages}
  {097402} (\bibinfo {year} {2012})}\BibitemShut {NoStop}%
\bibitem [{\citenamefont {Evlyukhin}\ \emph {et~al.}(2010)\citenamefont
  {Evlyukhin}, \citenamefont {Reinhardt}, \citenamefont {Seidel}, \citenamefont
  {Lukyanchuk},\ and\ \citenamefont {Chichkov}}]{Evlyukhin2010}%
  \BibitemOpen
  \bibfield  {author} {\bibinfo {author} {\bibfnamefont {A.~B.}\ \bibnamefont
  {Evlyukhin}}, \bibinfo {author} {\bibfnamefont {C.}~\bibnamefont
  {Reinhardt}}, \bibinfo {author} {\bibfnamefont {A.}~\bibnamefont {Seidel}},
  \bibinfo {author} {\bibfnamefont {B.~S.}\ \bibnamefont {Lukyanchuk}}, \ and\
  \bibinfo {author} {\bibfnamefont {B.~N.}\ \bibnamefont {Chichkov}},\
  }\href@noop {} {\bibfield  {journal} {\bibinfo  {journal} {Phys. Rev. B}\
  }\textbf {\bibinfo {volume} {82}},\ \bibinfo {pages} {45404} (\bibinfo {year}
  {2010})}\BibitemShut {NoStop}%
\bibitem [{\citenamefont {Moitra}\ \emph {et~al.}(2015)\citenamefont {Moitra},
  \citenamefont {Slovick}, \citenamefont {Li}, \citenamefont {Kravchencko},
  \citenamefont {Briggs}, \citenamefont {Krishnamurthy},\ and\ \citenamefont
  {Valentine}}]{Moitra2015}%
  \BibitemOpen
  \bibfield  {author} {\bibinfo {author} {\bibfnamefont {P.}~\bibnamefont
  {Moitra}}, \bibinfo {author} {\bibfnamefont {B.~A.}\ \bibnamefont {Slovick}},
  \bibinfo {author} {\bibfnamefont {W.}~\bibnamefont {Li}}, \bibinfo {author}
  {\bibfnamefont {I.~I.}\ \bibnamefont {Kravchencko}}, \bibinfo {author}
  {\bibfnamefont {D.~P.}\ \bibnamefont {Briggs}}, \bibinfo {author}
  {\bibfnamefont {S.}~\bibnamefont {Krishnamurthy}}, \ and\ \bibinfo {author}
  {\bibfnamefont {J.}~\bibnamefont {Valentine}},\ }\href {\doibase
  10.1021/acsphotonics.5b00148} {\bibfield  {journal} {\bibinfo  {journal} {ACS
  Photonics}\ }\textbf {\bibinfo {volume} {2}},\ \bibinfo {pages} {692}
  (\bibinfo {year} {2015})}\BibitemShut {NoStop}%
\bibitem [{\citenamefont {Arbabi}\ \emph {et~al.}(2015)\citenamefont {Arbabi},
  \citenamefont {Horie}, \citenamefont {Bagheri},\ and\ \citenamefont
  {Faraon}}]{Arbabi2015}%
  \BibitemOpen
  \bibfield  {author} {\bibinfo {author} {\bibfnamefont {A.}~\bibnamefont
  {Arbabi}}, \bibinfo {author} {\bibfnamefont {Y.}~\bibnamefont {Horie}},
  \bibinfo {author} {\bibfnamefont {M.}~\bibnamefont {Bagheri}}, \ and\
  \bibinfo {author} {\bibfnamefont {A.}~\bibnamefont {Faraon}},\ }\href@noop {}
  {\bibfield  {journal} {\bibinfo  {journal} {Nat. Nanotechnol.}\ }\textbf
  {\bibinfo {volume} {10}},\ \bibinfo {pages} {937} (\bibinfo {year}
  {2015})}\BibitemShut {NoStop}%
\bibitem [{\citenamefont {Shalaev}\ \emph {et~al.}(2015)\citenamefont
  {Shalaev}, \citenamefont {Sun}, \citenamefont {Tsukernik}, \citenamefont
  {Pandey}, \citenamefont {Nikolskiy},\ and\ \citenamefont
  {Litchinitser}}]{Shalaev2015}%
  \BibitemOpen
  \bibfield  {author} {\bibinfo {author} {\bibfnamefont {M.~I.}\ \bibnamefont
  {Shalaev}}, \bibinfo {author} {\bibfnamefont {J.}~\bibnamefont {Sun}},
  \bibinfo {author} {\bibfnamefont {A.}~\bibnamefont {Tsukernik}}, \bibinfo
  {author} {\bibfnamefont {A.}~\bibnamefont {Pandey}}, \bibinfo {author}
  {\bibfnamefont {K.}~\bibnamefont {Nikolskiy}}, \ and\ \bibinfo {author}
  {\bibfnamefont {N.~M.}\ \bibnamefont {Litchinitser}},\ }\href@noop {}
  {\bibfield  {journal} {\bibinfo  {journal} {Nano Lett.}\ }\textbf {\bibinfo
  {volume} {15}},\ \bibinfo {pages} {6261} (\bibinfo {year}
  {2015})}\BibitemShut {NoStop}%
\bibitem [{\citenamefont {Khorasaninejad}\ \emph {et~al.}(2015)\citenamefont
  {Khorasaninejad}, \citenamefont {Chen}, \citenamefont {Devlin}, \citenamefont
  {Oh}, \citenamefont {Zhu},\ and\ \citenamefont
  {Capasso}}]{Khorasaninejad2015}%
  \BibitemOpen
  \bibfield  {author} {\bibinfo {author} {\bibfnamefont {M.}~\bibnamefont
  {Khorasaninejad}}, \bibinfo {author} {\bibfnamefont {W.~T.}\ \bibnamefont
  {Chen}}, \bibinfo {author} {\bibfnamefont {R.~C.}\ \bibnamefont {Devlin}},
  \bibinfo {author} {\bibfnamefont {J.}~\bibnamefont {Oh}}, \bibinfo {author}
  {\bibfnamefont {A.~Y.}\ \bibnamefont {Zhu}}, \ and\ \bibinfo {author}
  {\bibfnamefont {F.}~\bibnamefont {Capasso}},\ }\href@noop {} {\bibfield
  {journal} {\bibinfo  {journal} {Science}\ }\textbf {\bibinfo {volume}
  {352}},\ \bibinfo {pages} {1190} (\bibinfo {year} {22015})}\BibitemShut
  {NoStop}%
\bibitem [{\citenamefont {Yu}\ \emph {et~al.}(2015)\citenamefont {Yu},
  \citenamefont {Zhu}, \citenamefont {Paniagua-Dom\'{\i}nguez}, \citenamefont
  {Fu}, \citenamefont {Luk'yanchuk},\ and\ \citenamefont {Kuznetsov}}]{Yu2015}%
  \BibitemOpen
  \bibfield  {author} {\bibinfo {author} {\bibfnamefont {Y.~F.}\ \bibnamefont
  {Yu}}, \bibinfo {author} {\bibfnamefont {A.~Y.}\ \bibnamefont {Zhu}},
  \bibinfo {author} {\bibfnamefont {R.}~\bibnamefont
  {Paniagua-Dom\'{\i}nguez}}, \bibinfo {author} {\bibfnamefont {Y.~H.}\
  \bibnamefont {Fu}}, \bibinfo {author} {\bibfnamefont {B.}~\bibnamefont
  {Luk'yanchuk}}, \ and\ \bibinfo {author} {\bibfnamefont {A.~I.}\ \bibnamefont
  {Kuznetsov}},\ }\href@noop {} {\bibfield  {journal} {\bibinfo  {journal}
  {Laser Photon. Rev.}\ }\textbf {\bibinfo {volume} {9}},\ \bibinfo {pages}
  {412} (\bibinfo {year} {2015})}\BibitemShut {NoStop}%
\bibitem [{\citenamefont {Rolly}\ \emph {et~al.}(2012)\citenamefont {Rolly},
  \citenamefont {Bebey}, \citenamefont {Bidault}, \citenamefont {Stout},\ and\
  \citenamefont {Bonod}}]{Bonod2012}%
  \BibitemOpen
  \bibfield  {author} {\bibinfo {author} {\bibfnamefont {B.}~\bibnamefont
  {Rolly}}, \bibinfo {author} {\bibfnamefont {B.}~\bibnamefont {Bebey}},
  \bibinfo {author} {\bibfnamefont {S.}~\bibnamefont {Bidault}}, \bibinfo
  {author} {\bibfnamefont {B.}~\bibnamefont {Stout}}, \ and\ \bibinfo {author}
  {\bibfnamefont {N.}~\bibnamefont {Bonod}},\ }\href {\doibase
  10.1103/PhysRevB.85.245432} {\bibfield  {journal} {\bibinfo  {journal} {Phys.
  Rev. B}\ }\textbf {\bibinfo {volume} {85}},\ \bibinfo {pages} {245432}
  (\bibinfo {year} {2012})}\BibitemShut {NoStop}%
\bibitem [{\citenamefont {Albella}\ \emph {et~al.}(2013)\citenamefont
  {Albella}, \citenamefont {Poyli}, \citenamefont {Schmidt}, \citenamefont
  {Maier}, \citenamefont {Moreno}, \citenamefont {Saenz},\ and\ \citenamefont
  {Aizpurua}}]{Albella2013}%
  \BibitemOpen
  \bibfield  {author} {\bibinfo {author} {\bibfnamefont {P.}~\bibnamefont
  {Albella}}, \bibinfo {author} {\bibfnamefont {M.~A.}\ \bibnamefont {Poyli}},
  \bibinfo {author} {\bibfnamefont {M.~K.}\ \bibnamefont {Schmidt}}, \bibinfo
  {author} {\bibfnamefont {S.~A.}\ \bibnamefont {Maier}}, \bibinfo {author}
  {\bibfnamefont {F.}~\bibnamefont {Moreno}}, \bibinfo {author} {\bibfnamefont
  {J.~J.}\ \bibnamefont {Saenz}}, \ and\ \bibinfo {author} {\bibfnamefont
  {J.}~\bibnamefont {Aizpurua}},\ }\href {\doibase 10.1021/jp4027018}
  {\bibfield  {journal} {\bibinfo  {journal} {J. Phys. Chem. C}\ }\textbf
  {\bibinfo {volume} {117}},\ \bibinfo {pages} {13573} (\bibinfo {year}
  {2013})}\BibitemShut {NoStop}%
\bibitem [{\citenamefont {Dmitriev}\ \emph {et~al.}(2016)\citenamefont
  {Dmitriev}, \citenamefont {Baranov}, \citenamefont {Milichko}, \citenamefont
  {Makarov}, \citenamefont {Mukhin}, \citenamefont {Samusev}, \citenamefont
  {Krasnok}, \citenamefont {Belov},\ and\ \citenamefont {Kivshar}}]{Baranov16}%
  \BibitemOpen
  \bibfield  {author} {\bibinfo {author} {\bibfnamefont {P.~A.}\ \bibnamefont
  {Dmitriev}}, \bibinfo {author} {\bibfnamefont {D.~G.}\ \bibnamefont
  {Baranov}}, \bibinfo {author} {\bibfnamefont {V.~A.}\ \bibnamefont
  {Milichko}}, \bibinfo {author} {\bibfnamefont {S.~V.}\ \bibnamefont
  {Makarov}}, \bibinfo {author} {\bibfnamefont {I.~S.}\ \bibnamefont {Mukhin}},
  \bibinfo {author} {\bibfnamefont {A.~K.}\ \bibnamefont {Samusev}}, \bibinfo
  {author} {\bibfnamefont {A.~E.}\ \bibnamefont {Krasnok}}, \bibinfo {author}
  {\bibfnamefont {P.~A.}\ \bibnamefont {Belov}}, \ and\ \bibinfo {author}
  {\bibfnamefont {Y.~S.}\ \bibnamefont {Kivshar}},\ }\href@noop {} {\bibfield
  {journal} {\bibinfo  {journal} {Nanoscale}\ }\textbf {\bibinfo {volume}
  {8}},\ \bibinfo {pages} {9721} (\bibinfo {year} {2016})}\BibitemShut
  {NoStop}%
\bibitem [{\citenamefont {Krasnok}\ \emph {et~al.}(2016)\citenamefont
  {Krasnok}, \citenamefont {Glybovski}, \citenamefont {Petrov}, \citenamefont
  {Makarov}, \citenamefont {Savelev}, \citenamefont {Belov}, \citenamefont
  {Simovski},\ and\ \citenamefont {Kivshar}}]{Krasnok2016}%
  \BibitemOpen
  \bibfield  {author} {\bibinfo {author} {\bibfnamefont {A.}~\bibnamefont
  {Krasnok}}, \bibinfo {author} {\bibfnamefont {S.}~\bibnamefont {Glybovski}},
  \bibinfo {author} {\bibfnamefont {M.}~\bibnamefont {Petrov}}, \bibinfo
  {author} {\bibfnamefont {S.}~\bibnamefont {Makarov}}, \bibinfo {author}
  {\bibfnamefont {R.}~\bibnamefont {Savelev}}, \bibinfo {author} {\bibfnamefont
  {P.}~\bibnamefont {Belov}}, \bibinfo {author} {\bibfnamefont
  {C.}~\bibnamefont {Simovski}}, \ and\ \bibinfo {author} {\bibfnamefont
  {Y.}~\bibnamefont {Kivshar}},\ }\href {\doibase 10.1063/1.4952740} {\bibfield
   {journal} {\bibinfo  {journal} {Appl. Phys. Lett.}\ }\textbf {\bibinfo
  {volume} {108}},\ \bibinfo {pages} {211105} (\bibinfo {year}
  {2016})}\BibitemShut {NoStop}%
\bibitem [{\citenamefont {Krasnok}\ \emph {et~al.}(2012)\citenamefont
  {Krasnok}, \citenamefont {Miroshnichenko}, \citenamefont {Belov},\ and\
  \citenamefont {Kivshar}}]{Krasnok2012}%
  \BibitemOpen
  \bibfield  {author} {\bibinfo {author} {\bibfnamefont {A.~E.}\ \bibnamefont
  {Krasnok}}, \bibinfo {author} {\bibfnamefont {A.~E.}\ \bibnamefont
  {Miroshnichenko}}, \bibinfo {author} {\bibfnamefont {P.~a.}\ \bibnamefont
  {Belov}}, \ and\ \bibinfo {author} {\bibfnamefont {Y.~S.}\ \bibnamefont
  {Kivshar}},\ }\href {http://www.ncbi.nlm.nih.gov/pubmed/23037107} {\bibfield
  {journal} {\bibinfo  {journal} {Opt. Express}\ }\textbf {\bibinfo {volume}
  {20}},\ \bibinfo {pages} {20599} (\bibinfo {year} {2012})}\BibitemShut
  {NoStop}%
\bibitem [{\citenamefont {Garc{\'{i}}a-Etxarri}\ \emph
  {et~al.}(2011)\citenamefont {Garc{\'{i}}a-Etxarri}, \citenamefont
  {G{\'{o}}mez-Medina}, \citenamefont {Froufe-P{\'{e}}rez}, \citenamefont
  {L{\'{o}}pez}, \citenamefont {Chantada}, \citenamefont {Scheffold},
  \citenamefont {Aizpurua}, \citenamefont {Nieto-Vesperinas},\ and\
  \citenamefont {S{\'{a}}enz}}]{Garcia-Etxarri2011}%
  \BibitemOpen
  \bibfield  {author} {\bibinfo {author} {\bibfnamefont {A.}~\bibnamefont
  {Garc{\'{i}}a-Etxarri}}, \bibinfo {author} {\bibfnamefont {R.}~\bibnamefont
  {G{\'{o}}mez-Medina}}, \bibinfo {author} {\bibfnamefont {L.~S.}\ \bibnamefont
  {Froufe-P{\'{e}}rez}}, \bibinfo {author} {\bibfnamefont {C.}~\bibnamefont
  {L{\'{o}}pez}}, \bibinfo {author} {\bibfnamefont {L.}~\bibnamefont
  {Chantada}}, \bibinfo {author} {\bibfnamefont {F.}~\bibnamefont {Scheffold}},
  \bibinfo {author} {\bibfnamefont {J.}~\bibnamefont {Aizpurua}}, \bibinfo
  {author} {\bibfnamefont {M.}~\bibnamefont {Nieto-Vesperinas}}, \ and\
  \bibinfo {author} {\bibfnamefont {J.~J.}\ \bibnamefont {S{\'{a}}enz}},\
  }\href@noop {} {\bibfield  {journal} {\bibinfo  {journal} {Opt. Express}\
  }\textbf {\bibinfo {volume} {19}},\ \bibinfo {pages} {4815} (\bibinfo {year}
  {2011})}\BibitemShut {NoStop}%
\bibitem [{\citenamefont {Geffrin}\ \emph {et~al.}(2012)\citenamefont
  {Geffrin}, \citenamefont {Garc\'{\i}a-C\'{a}mara}, \citenamefont
  {G\'{o}mez-Medina}, \citenamefont {Albella}, \citenamefont
  {Froufe-P\'{e}rez}, \citenamefont {Eyraud}, \citenamefont {Litman},
  \citenamefont {Vaillon}, \citenamefont {Gonz\'{a}lez}, \citenamefont
  {Nieto-Vesperinas}, \citenamefont {S\'{a}enz},\ and\ \citenamefont
  {Moreno}}]{Geffrin2012}%
  \BibitemOpen
  \bibfield  {author} {\bibinfo {author} {\bibfnamefont {J.~M.}\ \bibnamefont
  {Geffrin}}, \bibinfo {author} {\bibfnamefont {B.}~\bibnamefont
  {Garc\'{\i}a-C\'{a}mara}}, \bibinfo {author} {\bibfnamefont {R.}~\bibnamefont
  {G\'{o}mez-Medina}}, \bibinfo {author} {\bibfnamefont {P.}~\bibnamefont
  {Albella}}, \bibinfo {author} {\bibfnamefont {L.~S.}\ \bibnamefont
  {Froufe-P\'{e}rez}}, \bibinfo {author} {\bibfnamefont {C.}~\bibnamefont
  {Eyraud}}, \bibinfo {author} {\bibfnamefont {A.}~\bibnamefont {Litman}},
  \bibinfo {author} {\bibfnamefont {R.}~\bibnamefont {Vaillon}}, \bibinfo
  {author} {\bibfnamefont {F.}~\bibnamefont {Gonz\'{a}lez}}, \bibinfo {author}
  {\bibfnamefont {M.}~\bibnamefont {Nieto-Vesperinas}}, \bibinfo {author}
  {\bibfnamefont {J.~J.}\ \bibnamefont {S\'{a}enz}}, \ and\ \bibinfo {author}
  {\bibfnamefont {F.}~\bibnamefont {Moreno}},\ }\href {\doibase
  10.1038/ncomms2167} {\bibfield  {journal} {\bibinfo  {journal} {Nat.
  Commun.}\ }\textbf {\bibinfo {volume} {3}},\ \bibinfo {pages} {1171}
  (\bibinfo {year} {2012})}\BibitemShut {NoStop}%
\bibitem [{\citenamefont {Evlyukhin}\ \emph {et~al.}(2014)\citenamefont
  {Evlyukhin}, \citenamefont {Eriksen}, \citenamefont {Cheng}, \citenamefont
  {Beermann}, \citenamefont {Reinhardt}, \citenamefont {Petrov}, \citenamefont
  {Prorok}, \citenamefont {Eich}, \citenamefont {Chichkov},\ and\ \citenamefont
  {Bozhevolnyi}}]{Evlyukhin2014}%
  \BibitemOpen
  \bibfield  {author} {\bibinfo {author} {\bibfnamefont {A.~B.}\ \bibnamefont
  {Evlyukhin}}, \bibinfo {author} {\bibfnamefont {R.~L.}\ \bibnamefont
  {Eriksen}}, \bibinfo {author} {\bibfnamefont {W.}~\bibnamefont {Cheng}},
  \bibinfo {author} {\bibfnamefont {J.}~\bibnamefont {Beermann}}, \bibinfo
  {author} {\bibfnamefont {C.}~\bibnamefont {Reinhardt}}, \bibinfo {author}
  {\bibfnamefont {A.}~\bibnamefont {Petrov}}, \bibinfo {author} {\bibfnamefont
  {S.}~\bibnamefont {Prorok}}, \bibinfo {author} {\bibfnamefont
  {M.}~\bibnamefont {Eich}}, \bibinfo {author} {\bibfnamefont {B.~N.}\
  \bibnamefont {Chichkov}}, \ and\ \bibinfo {author} {\bibfnamefont {S.~I.}\
  \bibnamefont {Bozhevolnyi}},\ }\href@noop {} {\bibfield  {journal} {\bibinfo
  {journal} {Sci. Rep.}\ }\textbf {\bibinfo {volume} {4}},\ \bibinfo {pages}
  {4126} (\bibinfo {year} {2014})}\BibitemShut {NoStop}%
\bibitem [{\citenamefont {Albella}\ \emph {et~al.}(2015)\citenamefont
  {Albella}, \citenamefont {Shibanuma},\ and\ \citenamefont
  {Maier}}]{Albella2015}%
  \BibitemOpen
  \bibfield  {author} {\bibinfo {author} {\bibfnamefont {P.}~\bibnamefont
  {Albella}}, \bibinfo {author} {\bibfnamefont {T.}~\bibnamefont {Shibanuma}},
  \ and\ \bibinfo {author} {\bibfnamefont {S.~A.}\ \bibnamefont {Maier}},\
  }\href {\doibase 10.1038/srep18322} {\bibfield  {journal} {\bibinfo
  {journal} {Sci. Rep.}\ }\textbf {\bibinfo {volume} {5}},\ \bibinfo {pages}
  {18322} (\bibinfo {year} {2015})}\BibitemShut {NoStop}%
\bibitem [{\citenamefont {Fu}\ \emph {et~al.}(2013)\citenamefont {Fu},
  \citenamefont {Kuznetsov}, \citenamefont {Miroshnichenko}, \citenamefont
  {Yu},\ and\ \citenamefont {Luk'yanchuk}}]{Fu2013}%
  \BibitemOpen
  \bibfield  {author} {\bibinfo {author} {\bibfnamefont {Y.~H.}\ \bibnamefont
  {Fu}}, \bibinfo {author} {\bibfnamefont {A.~I.}\ \bibnamefont {Kuznetsov}},
  \bibinfo {author} {\bibfnamefont {A.~E.}\ \bibnamefont {Miroshnichenko}},
  \bibinfo {author} {\bibfnamefont {Y.~F.}\ \bibnamefont {Yu}}, \ and\ \bibinfo
  {author} {\bibfnamefont {B.}~\bibnamefont {Luk'yanchuk}},\ }\href {\doibase
  10.1038/ncomms2538} {\bibfield  {journal} {\bibinfo  {journal} {Nat.
  Commun.}\ }\textbf {\bibinfo {volume} {4}},\ \bibinfo {pages} {1527}
  (\bibinfo {year} {2013})}\BibitemShut {NoStop}%
\bibitem [{\citenamefont {Savelev}\ \emph {et~al.}(2015)\citenamefont
  {Savelev}, \citenamefont {Makarov}, \citenamefont {Krasnok},\ and\
  \citenamefont {Belov}}]{Savelev2015}%
  \BibitemOpen
  \bibfield  {author} {\bibinfo {author} {\bibfnamefont {R.~S.}\ \bibnamefont
  {Savelev}}, \bibinfo {author} {\bibfnamefont {S.~V.}\ \bibnamefont
  {Makarov}}, \bibinfo {author} {\bibfnamefont {A.~E.}\ \bibnamefont
  {Krasnok}}, \ and\ \bibinfo {author} {\bibfnamefont {P.~A.}\ \bibnamefont
  {Belov}},\ }\href {\doibase 10.1134/S0030400X15100240} {\bibfield  {journal}
  {\bibinfo  {journal} {Opt. Spectrosc.}\ }\textbf {\bibinfo {volume} {119}},\
  \bibinfo {pages} {551} (\bibinfo {year} {2015})}\BibitemShut {NoStop}%
\bibitem [{\citenamefont {Li}\ \emph {et~al.}(2015)\citenamefont {Li},
  \citenamefont {Baranov}, \citenamefont {Krasnok},\ and\ \citenamefont
  {Belov}}]{Li2015}%
  \BibitemOpen
  \bibfield  {author} {\bibinfo {author} {\bibfnamefont {S.~V.}\ \bibnamefont
  {Li}}, \bibinfo {author} {\bibfnamefont {D.~G.}\ \bibnamefont {Baranov}},
  \bibinfo {author} {\bibfnamefont {A.~E.}\ \bibnamefont {Krasnok}}, \ and\
  \bibinfo {author} {\bibfnamefont {P.~A.}\ \bibnamefont {Belov}},\ }\href
  {\doibase 10.1063/1.4934757} {\bibfield  {journal} {\bibinfo  {journal}
  {Appl. Phys. Lett.}\ }\textbf {\bibinfo {volume} {107}},\ \bibinfo {pages}
  {171101} (\bibinfo {year} {2015})}\BibitemShut {NoStop}%
\bibitem [{\citenamefont {Markovich}\ \emph {et~al.}(2016)\citenamefont
  {Markovich}, \citenamefont {Baryshnikova}, \citenamefont {Shalin},
  \citenamefont {Samusev}, \citenamefont {Krasnok},\ and\ \citenamefont
  {Belov}}]{Markovich2016}%
  \BibitemOpen
  \bibfield  {author} {\bibinfo {author} {\bibfnamefont {D.}~\bibnamefont
  {Markovich}}, \bibinfo {author} {\bibfnamefont {K.}~\bibnamefont
  {Baryshnikova}}, \bibinfo {author} {\bibfnamefont {A.}~\bibnamefont
  {Shalin}}, \bibinfo {author} {\bibfnamefont {A.}~\bibnamefont {Samusev}},
  \bibinfo {author} {\bibfnamefont {A.}~\bibnamefont {Krasnok}}, \ and\
  \bibinfo {author} {\bibfnamefont {P.}~\bibnamefont {Belov}},\ }\href
  {http://www.nature.com/articles/srep22546} {\bibfield  {journal} {\bibinfo
  {journal} {Sci. Rep.}\ }\textbf {\bibinfo {volume} {6}},\ \bibinfo {pages}
  {22546} (\bibinfo {year} {2016})}\BibitemShut {NoStop}%
\bibitem [{\citenamefont {Shcherbakov}\ \emph {et~al.}(2014)\citenamefont
  {Shcherbakov}, \citenamefont {Neshev}, \citenamefont {Hopkins}, \citenamefont
  {Shorokhov}, \citenamefont {Staude}, \citenamefont {Melik-Gaykazyan},
  \citenamefont {Decker}, \citenamefont {Ezhov}, \citenamefont
  {Miroshnichenko}, \citenamefont {Brener}, \citenamefont {Fedyanin},\ and\
  \citenamefont {Kivshar}}]{Shcherbakov2014}%
  \BibitemOpen
  \bibfield  {author} {\bibinfo {author} {\bibfnamefont {M.~R.}\ \bibnamefont
  {Shcherbakov}}, \bibinfo {author} {\bibfnamefont {D.~N.}\ \bibnamefont
  {Neshev}}, \bibinfo {author} {\bibfnamefont {B.}~\bibnamefont {Hopkins}},
  \bibinfo {author} {\bibfnamefont {A.~S.}\ \bibnamefont {Shorokhov}}, \bibinfo
  {author} {\bibfnamefont {I.}~\bibnamefont {Staude}}, \bibinfo {author}
  {\bibfnamefont {E.~V.}\ \bibnamefont {Melik-Gaykazyan}}, \bibinfo {author}
  {\bibfnamefont {M.}~\bibnamefont {Decker}}, \bibinfo {author} {\bibfnamefont
  {A.~A.}\ \bibnamefont {Ezhov}}, \bibinfo {author} {\bibfnamefont {A.~E.}\
  \bibnamefont {Miroshnichenko}}, \bibinfo {author} {\bibfnamefont
  {I.}~\bibnamefont {Brener}}, \bibinfo {author} {\bibfnamefont {A.~A.}\
  \bibnamefont {Fedyanin}}, \ and\ \bibinfo {author} {\bibfnamefont {Y.~S.}\
  \bibnamefont {Kivshar}},\ }\href {\doibase 10.1021/nl503029j} {\bibfield
  {journal} {\bibinfo  {journal} {Nano Lett.}\ }\textbf {\bibinfo {volume}
  {14}},\ \bibinfo {pages} {6488} (\bibinfo {year} {2014})}\BibitemShut
  {NoStop}%
\bibitem [{\citenamefont {Makarov}\ \emph {et~al.}(2015)\citenamefont
  {Makarov}, \citenamefont {Kudryashov}, \citenamefont {Mukhin}, \citenamefont
  {Mozharov}, \citenamefont {Milichko}, \citenamefont {Krasnok},\ and\
  \citenamefont {Belov}}]{Makarov2015}%
  \BibitemOpen
  \bibfield  {author} {\bibinfo {author} {\bibfnamefont {S.}~\bibnamefont
  {Makarov}}, \bibinfo {author} {\bibfnamefont {S.}~\bibnamefont {Kudryashov}},
  \bibinfo {author} {\bibfnamefont {I.}~\bibnamefont {Mukhin}}, \bibinfo
  {author} {\bibfnamefont {A.}~\bibnamefont {Mozharov}}, \bibinfo {author}
  {\bibfnamefont {V.}~\bibnamefont {Milichko}}, \bibinfo {author}
  {\bibfnamefont {A.}~\bibnamefont {Krasnok}}, \ and\ \bibinfo {author}
  {\bibfnamefont {P.}~\bibnamefont {Belov}},\ }\href@noop {} {\bibfield
  {journal} {\bibinfo  {journal} {Nano Lett.}\ }\textbf {\bibinfo {volume}
  {15}},\ \bibinfo {pages} {6187} (\bibinfo {year} {2015})}\BibitemShut
  {NoStop}%
\bibitem [{\citenamefont {Shcherbakov}\ \emph {et~al.}(2015)\citenamefont
  {Shcherbakov}, \citenamefont {Vabishchevich}, \citenamefont {Shorokhov},
  \citenamefont {Chong}, \citenamefont {Choi}, \citenamefont {Staude},
  \citenamefont {Miroshnichenko}, \citenamefont {Neshev}, \citenamefont
  {Fedyanin},\ and\ \citenamefont {Kivshar}}]{Shcherbakov2015}%
  \BibitemOpen
  \bibfield  {author} {\bibinfo {author} {\bibfnamefont {M.~R.}\ \bibnamefont
  {Shcherbakov}}, \bibinfo {author} {\bibfnamefont {P.~P.}\ \bibnamefont
  {Vabishchevich}}, \bibinfo {author} {\bibfnamefont {A.~S.}\ \bibnamefont
  {Shorokhov}}, \bibinfo {author} {\bibfnamefont {K.~E.}\ \bibnamefont
  {Chong}}, \bibinfo {author} {\bibfnamefont {D.-Y.}\ \bibnamefont {Choi}},
  \bibinfo {author} {\bibfnamefont {I.}~\bibnamefont {Staude}}, \bibinfo
  {author} {\bibfnamefont {A.~E.}\ \bibnamefont {Miroshnichenko}}, \bibinfo
  {author} {\bibfnamefont {D.~N.}\ \bibnamefont {Neshev}}, \bibinfo {author}
  {\bibfnamefont {A.~A.}\ \bibnamefont {Fedyanin}}, \ and\ \bibinfo {author}
  {\bibfnamefont {Y.~S.}\ \bibnamefont {Kivshar}},\ }\href {\doibase
  10.1021/acs.nanolett.5b02989} {\bibfield  {journal} {\bibinfo  {journal}
  {Nano Lett.}\ }\textbf {\bibinfo {volume} {15}},\ \bibinfo {pages} {6985}
  (\bibinfo {year} {2015})}\BibitemShut {NoStop}%
\bibitem [{\citenamefont {Baranov}\ \emph {et~al.}(2016)\citenamefont
  {Baranov}, \citenamefont {Makarov}, \citenamefont {Milichko}, \citenamefont
  {Kudryashov}, \citenamefont {Krasnok},\ and\ \citenamefont {Belov}}]{Plasma}%
  \BibitemOpen
  \bibfield  {author} {\bibinfo {author} {\bibfnamefont {D.~G.}\ \bibnamefont
  {Baranov}}, \bibinfo {author} {\bibfnamefont {S.~V.}\ \bibnamefont
  {Makarov}}, \bibinfo {author} {\bibfnamefont {V.~A.}\ \bibnamefont
  {Milichko}}, \bibinfo {author} {\bibfnamefont {S.~I.}\ \bibnamefont
  {Kudryashov}}, \bibinfo {author} {\bibfnamefont {A.~E.}\ \bibnamefont
  {Krasnok}}, \ and\ \bibinfo {author} {\bibfnamefont {P.~A.}\ \bibnamefont
  {Belov}},\ }\href {\doibase 10.1021/acsphotonics.6b00358} {\bibfield
  {journal} {\bibinfo  {journal} {ACS Photonics}\ }\textbf {\bibinfo {volume}
  {3}},\ \bibinfo {pages} {1546} (\bibinfo {year} {2016})}\BibitemShut
  {NoStop}%
\bibitem [{\citenamefont {Maier}(2006)}]{Maier06}%
  \BibitemOpen
  \bibfield  {author} {\bibinfo {author} {\bibfnamefont {S.~A.}\ \bibnamefont
  {Maier}},\ }\href@noop {} {\bibfield  {journal} {\bibinfo  {journal} {Opt.
  Express}\ }\textbf {\bibinfo {volume} {14}},\ \bibinfo {pages} {1957}
  (\bibinfo {year} {2006})}\BibitemShut {NoStop}%
\bibitem [{\citenamefont {Bharadwaj}\ \emph {et~al.}(2009)\citenamefont
  {Bharadwaj}, \citenamefont {Deutsch},\ and\ \citenamefont
  {Novotny}}]{NovotnyAOP}%
  \BibitemOpen
  \bibfield  {author} {\bibinfo {author} {\bibfnamefont {P.}~\bibnamefont
  {Bharadwaj}}, \bibinfo {author} {\bibfnamefont {B.}~\bibnamefont {Deutsch}},
  \ and\ \bibinfo {author} {\bibfnamefont {L.}~\bibnamefont {Novotny}},\
  }\href@noop {} {\bibfield  {journal} {\bibinfo  {journal} {Adv. Opt.
  Photon.}\ }\textbf {\bibinfo {volume} {1}},\ \bibinfo {pages} {438} (\bibinfo
  {year} {2009})}\BibitemShut {NoStop}%
\bibitem [{\citenamefont {Agio}(2012)}]{Agio2012}%
  \BibitemOpen
  \bibfield  {author} {\bibinfo {author} {\bibfnamefont {M.}~\bibnamefont
  {Agio}},\ }\href {\doibase 10.1039/C1NR11116G} {\bibfield  {journal}
  {\bibinfo  {journal} {Nanoscale}\ }\textbf {\bibinfo {volume} {4}},\ \bibinfo
  {pages} {692} (\bibinfo {year} {2012})}\BibitemShut {NoStop}%
\bibitem [{\citenamefont {Seok}\ \emph {et~al.}(2011)\citenamefont {Seok},
  \citenamefont {Jamshidi}, \citenamefont {Kim}, \citenamefont {Dhuey},
  \citenamefont {Lakhani}, \citenamefont {Choo}, \citenamefont {Schuck},
  \citenamefont {Cabrini}, \citenamefont {Schwartzberg}, \citenamefont {Bokor},
  \citenamefont {Yablonovitch},\ and\ \citenamefont {Wu}}]{Seok11}%
  \BibitemOpen
  \bibfield  {author} {\bibinfo {author} {\bibfnamefont {T.~J.}\ \bibnamefont
  {Seok}}, \bibinfo {author} {\bibfnamefont {A.}~\bibnamefont {Jamshidi}},
  \bibinfo {author} {\bibfnamefont {M.}~\bibnamefont {Kim}}, \bibinfo {author}
  {\bibfnamefont {S.}~\bibnamefont {Dhuey}}, \bibinfo {author} {\bibfnamefont
  {A.}~\bibnamefont {Lakhani}}, \bibinfo {author} {\bibfnamefont
  {H.}~\bibnamefont {Choo}}, \bibinfo {author} {\bibfnamefont {P.~J.}\
  \bibnamefont {Schuck}}, \bibinfo {author} {\bibfnamefont {S.}~\bibnamefont
  {Cabrini}}, \bibinfo {author} {\bibfnamefont {A.~M.}\ \bibnamefont
  {Schwartzberg}}, \bibinfo {author} {\bibfnamefont {J.}~\bibnamefont {Bokor}},
  \bibinfo {author} {\bibfnamefont {E.}~\bibnamefont {Yablonovitch}}, \ and\
  \bibinfo {author} {\bibfnamefont {M.~C.}\ \bibnamefont {Wu}},\ }\href@noop {}
  {\bibfield  {journal} {\bibinfo  {journal} {Nano Letters}\ }\textbf {\bibinfo
  {volume} {11}},\ \bibinfo {pages} {2606} (\bibinfo {year}
  {2011})}\BibitemShut {NoStop}%
\bibitem [{\citenamefont {Klingshirn}(2012)}]{Klingshirn}%
  \BibitemOpen
  \bibfield  {author} {\bibinfo {author} {\bibfnamefont {C.~F.}\ \bibnamefont
  {Klingshirn}},\ }\href@noop {} {\emph {\bibinfo {title} {Semiconductor
  Optics}}}\ (\bibinfo  {publisher} {Springer},\ \bibinfo {year}
  {2012})\BibitemShut {NoStop}%
\bibitem [{\citenamefont {Yu}\ and\ \citenamefont {Cardona}(2010)}]{Cardona}%
  \BibitemOpen
  \bibfield  {author} {\bibinfo {author} {\bibfnamefont {P.~Y.}\ \bibnamefont
  {Yu}}\ and\ \bibinfo {author} {\bibfnamefont {M.}~\bibnamefont {Cardona}},\
  }\href@noop {} {\emph {\bibinfo {title} {Fundamentals of Semiconductors}}}\
  (\bibinfo  {publisher} {Springer},\ \bibinfo {year} {2010})\BibitemShut
  {NoStop}%
\bibitem [{\citenamefont {Nussenzveig}(1972)}]{Nussenzveig}%
  \BibitemOpen
  \bibfield  {author} {\bibinfo {author} {\bibfnamefont {H.~M.}\ \bibnamefont
  {Nussenzveig}},\ }\href@noop {} {\emph {\bibinfo {title} {{Causality and
  Dispersion Relations}}}}\ (\bibinfo  {publisher} {Academic Press},\ \bibinfo
  {address} {New York},\ \bibinfo {year} {1972})\BibitemShut {NoStop}%
\bibitem [{\citenamefont {Caldwell}\ \emph {et~al.}(2015)\citenamefont
  {Caldwell}, \citenamefont {Lindsay}, \citenamefont {Giannini}, \citenamefont
  {Vurgaftman}, \citenamefont {Reinecke}, \citenamefont {Maier},\ and\
  \citenamefont {Glembocki}}]{CaldwellNanoph}%
  \BibitemOpen
  \bibfield  {author} {\bibinfo {author} {\bibfnamefont {J.~D.}\ \bibnamefont
  {Caldwell}}, \bibinfo {author} {\bibfnamefont {L.}~\bibnamefont {Lindsay}},
  \bibinfo {author} {\bibfnamefont {V.}~\bibnamefont {Giannini}}, \bibinfo
  {author} {\bibfnamefont {I.}~\bibnamefont {Vurgaftman}}, \bibinfo {author}
  {\bibfnamefont {T.~L.}\ \bibnamefont {Reinecke}}, \bibinfo {author}
  {\bibfnamefont {S.~A.}\ \bibnamefont {Maier}}, \ and\ \bibinfo {author}
  {\bibfnamefont {O.~J.}\ \bibnamefont {Glembocki}},\ }\href {\doibase
  10.1515/nanoph-2014-0003} {\bibfield  {journal} {\bibinfo  {journal}
  {Nanophotonics}\ }\textbf {\bibinfo {volume} {4}},\ \bibinfo {pages} {44}
  (\bibinfo {year} {2015})}\BibitemShut {NoStop}%
\bibitem [{\citenamefont {Green}\ and\ \citenamefont
  {Keevers}(1995)}]{Green1995}%
  \BibitemOpen
  \bibfield  {author} {\bibinfo {author} {\bibfnamefont {M.~A.}\ \bibnamefont
  {Green}}\ and\ \bibinfo {author} {\bibfnamefont {M.~J.}\ \bibnamefont
  {Keevers}},\ }\href@noop {} {\bibfield  {journal} {\bibinfo  {journal} {Prog.
  Photovoltaics Res. Appl.}\ }\textbf {\bibinfo {volume} {3}},\ \bibinfo
  {pages} {189} (\bibinfo {year} {1995})}\BibitemShut {NoStop}%
\bibitem [{\citenamefont {Li}(1980)}]{Li1980}%
  \BibitemOpen
  \bibfield  {author} {\bibinfo {author} {\bibfnamefont {H.}~\bibnamefont
  {Li}},\ }\href@noop {} {\bibfield  {journal} {\bibinfo  {journal} {J. Phys.
  Chem. Ref. Data}\ }\textbf {\bibinfo {volume} {9}},\ \bibinfo {pages} {561}
  (\bibinfo {year} {1980})}\BibitemShut {NoStop}%
\bibitem [{\citenamefont {Pierce}\ and\ \citenamefont
  {Spicer}(1972)}]{Pierce1972}%
  \BibitemOpen
  \bibfield  {author} {\bibinfo {author} {\bibfnamefont {D.}~\bibnamefont
  {Pierce}}\ and\ \bibinfo {author} {\bibfnamefont {W.}~\bibnamefont
  {Spicer}},\ }\href@noop {} {\bibfield  {journal} {\bibinfo  {journal} {Phys.
  Rev. B}\ }\textbf {\bibinfo {volume} {5}},\ \bibinfo {pages} {3017} (\bibinfo
  {year} {1972})}\BibitemShut {NoStop}%
\bibitem [{\citenamefont {Jellison}(1992)}]{Jellison1992}%
  \BibitemOpen
  \bibfield  {author} {\bibinfo {author} {\bibfnamefont {G.}~\bibnamefont
  {Jellison}},\ }\href@noop {} {\bibfield  {journal} {\bibinfo  {journal} {Opt.
  Mater.}\ }\textbf {\bibinfo {volume} {1}},\ \bibinfo {pages} {151} (\bibinfo
  {year} {1992})}\BibitemShut {NoStop}%
\bibitem [{\citenamefont {Aspnes}\ and\ \citenamefont
  {Studna}(1983)}]{Aspnes1983}%
  \BibitemOpen
  \bibfield  {author} {\bibinfo {author} {\bibfnamefont {D.}~\bibnamefont
  {Aspnes}}\ and\ \bibinfo {author} {\bibfnamefont {A.}~\bibnamefont
  {Studna}},\ }\href@noop {} {\bibfield  {journal} {\bibinfo  {journal} {Phys.
  Rev. B}\ }\textbf {\bibinfo {volume} {27}},\ \bibinfo {pages} {985} (\bibinfo
  {year} {1983})}\BibitemShut {NoStop}%
\bibitem [{\citenamefont {DeVore}(1951)}]{Devore1951}%
  \BibitemOpen
  \bibfield  {author} {\bibinfo {author} {\bibfnamefont {J.~R.}\ \bibnamefont
  {DeVore}},\ }\href@noop {} {\bibfield  {journal} {\bibinfo  {journal} {J.
  Opt. Soc. Am.}\ }\textbf {\bibinfo {volume} {41}},\ \bibinfo {pages} {416}
  (\bibinfo {year} {1951})}\BibitemShut {NoStop}%
\bibitem [{\citenamefont {Palik}(1998)}]{Palik}%
  \BibitemOpen
  \bibfield  {author} {\bibinfo {author} {\bibfnamefont {E.~D.}\ \bibnamefont
  {Palik}},\ }\href@noop {} {\emph {\bibinfo {title} {{Handbook of optical
  constants of solids}}}}\ (\bibinfo  {publisher} {Academic Press},\ \bibinfo
  {year} {1998})\BibitemShut {NoStop}%
\bibitem [{\citenamefont {Ferrini}\ \emph {et~al.}(1998)\citenamefont
  {Ferrini}, \citenamefont {Patrini},\ and\ \citenamefont
  {Franchi}}]{Ferrini1998}%
  \BibitemOpen
  \bibfield  {author} {\bibinfo {author} {\bibfnamefont {R.}~\bibnamefont
  {Ferrini}}, \bibinfo {author} {\bibfnamefont {M.}~\bibnamefont {Patrini}}, \
  and\ \bibinfo {author} {\bibfnamefont {S.}~\bibnamefont {Franchi}},\
  }\href@noop {} {\bibfield  {journal} {\bibinfo  {journal} {J. Appl. Phys.}\
  }\textbf {\bibinfo {volume} {84}},\ \bibinfo {pages} {4517} (\bibinfo {year}
  {1998})}\BibitemShut {NoStop}%
\bibitem [{\citenamefont {Caldwell}\ and\ \citenamefont
  {Fan}(1959)}]{Caldwell1959}%
  \BibitemOpen
  \bibfield  {author} {\bibinfo {author} {\bibfnamefont {R.~S.}\ \bibnamefont
  {Caldwell}}\ and\ \bibinfo {author} {\bibfnamefont {H.}~\bibnamefont {Fan}},\
  }\href@noop {} {\bibfield  {journal} {\bibinfo  {journal} {Phys. Rev.}\
  }\textbf {\bibinfo {volume} {114}},\ \bibinfo {pages} {664} (\bibinfo {year}
  {1959})}\BibitemShut {NoStop}%
\bibitem [{\citenamefont {Weiting}\ and\ \citenamefont
  {Yixun}(1990)}]{Weiting1990}%
  \BibitemOpen
  \bibfield  {author} {\bibinfo {author} {\bibfnamefont {F.}~\bibnamefont
  {Weiting}}\ and\ \bibinfo {author} {\bibfnamefont {Y.}~\bibnamefont
  {Yixun}},\ }\href@noop {} {\bibfield  {journal} {\bibinfo  {journal}
  {Infrared Phys.}\ }\textbf {\bibinfo {volume} {30}},\ \bibinfo {pages} {371}
  (\bibinfo {year} {1990})}\BibitemShut {NoStop}%
\bibitem [{\citenamefont {Okoye}(2002)}]{Okoye2002}%
  \BibitemOpen
  \bibfield  {author} {\bibinfo {author} {\bibfnamefont {C.}~\bibnamefont
  {Okoye}},\ }\href@noop {} {\bibfield  {journal} {\bibinfo  {journal} {J.
  Phys. Condens. Matter}\ }\textbf {\bibinfo {volume} {14}},\ \bibinfo {pages}
  {8625} (\bibinfo {year} {2002})}\BibitemShut {NoStop}%
\bibitem [{\citenamefont {Larruquert}\ \emph {et~al.}(2011)\citenamefont
  {Larruquert}, \citenamefont {P{\'e}rez-Mar{\'\i}n}, \citenamefont
  {Garc{\'\i}a-Cort{\'e}s}, \citenamefont {Rodr{\'\i}guez-de Marcos},
  \citenamefont {Azn{\'a}rez},\ and\ \citenamefont
  {M{\'e}ndez}}]{Larruquert2011}%
  \BibitemOpen
  \bibfield  {author} {\bibinfo {author} {\bibfnamefont {J.~I.}\ \bibnamefont
  {Larruquert}}, \bibinfo {author} {\bibfnamefont {A.~P.}\ \bibnamefont
  {P{\'e}rez-Mar{\'\i}n}}, \bibinfo {author} {\bibfnamefont {S.}~\bibnamefont
  {Garc{\'\i}a-Cort{\'e}s}}, \bibinfo {author} {\bibfnamefont {L.}~\bibnamefont
  {Rodr{\'\i}guez-de Marcos}}, \bibinfo {author} {\bibfnamefont {J.~A.}\
  \bibnamefont {Azn{\'a}rez}}, \ and\ \bibinfo {author} {\bibfnamefont {J.~A.}\
  \bibnamefont {M{\'e}ndez}},\ }\href@noop {} {\bibfield  {journal} {\bibinfo
  {journal} {J. Opt. Soc. Am. A}\ }\textbf {\bibinfo {volume} {28}},\ \bibinfo
  {pages} {2340} (\bibinfo {year} {2011})}\BibitemShut {NoStop}%
\bibitem [{\citenamefont {Moss}(1985)}]{Moss}%
  \BibitemOpen
  \bibfield  {author} {\bibinfo {author} {\bibfnamefont {T.~S.}\ \bibnamefont
  {Moss}},\ }\href@noop {} {\bibfield  {journal} {\bibinfo  {journal} {Phys.
  Status Solidi B}\ }\textbf {\bibinfo {volume} {131}},\ \bibinfo {pages} {415}
  (\bibinfo {year} {1985})}\BibitemShut {NoStop}%
\bibitem [{\citenamefont {Ravindra}\ \emph {et~al.}(2006)\citenamefont
  {Ravindra}, \citenamefont {Ganapathy},\ and\ \citenamefont
  {Choi}}]{Ravindra}%
  \BibitemOpen
  \bibfield  {author} {\bibinfo {author} {\bibfnamefont {N.}~\bibnamefont
  {Ravindra}}, \bibinfo {author} {\bibfnamefont {P.}~\bibnamefont {Ganapathy}},
  \ and\ \bibinfo {author} {\bibfnamefont {J.}~\bibnamefont {Choi}},\
  }\href@noop {} {\bibfield  {journal} {\bibinfo  {journal} {Infrared Phys.
  Technol.}\ }\textbf {\bibinfo {volume} {50}},\ \bibinfo {pages} {21}
  (\bibinfo {year} {2006})}\BibitemShut {NoStop}%
\bibitem [{\citenamefont {Herve}\ and\ \citenamefont
  {Vandamme}(1994)}]{Tripathy}%
  \BibitemOpen
  \bibfield  {author} {\bibinfo {author} {\bibfnamefont {P.}~\bibnamefont
  {Herve}}\ and\ \bibinfo {author} {\bibfnamefont {L.}~\bibnamefont
  {Vandamme}},\ }\href@noop {} {\bibfield  {journal} {\bibinfo  {journal}
  {Infrared Phys. Technol.}\ }\textbf {\bibinfo {volume} {35}},\ \bibinfo
  {pages} {609} (\bibinfo {year} {1994})}\BibitemShut {NoStop}%
\bibitem [{\citenamefont {Zywietz}\ \emph {et~al.}(2015)\citenamefont
  {Zywietz}, \citenamefont {Schmidt}, \citenamefont {Evlyukhin}, \citenamefont
  {Reinhardt}, \citenamefont {Aizpurua},\ and\ \citenamefont
  {Chichkov}}]{ZywietzACSPhoton2015}%
  \BibitemOpen
  \bibfield  {author} {\bibinfo {author} {\bibfnamefont {U.}~\bibnamefont
  {Zywietz}}, \bibinfo {author} {\bibfnamefont {M.}~\bibnamefont {Schmidt}},
  \bibinfo {author} {\bibfnamefont {A.}~\bibnamefont {Evlyukhin}}, \bibinfo
  {author} {\bibfnamefont {C.}~\bibnamefont {Reinhardt}}, \bibinfo {author}
  {\bibfnamefont {J.}~\bibnamefont {Aizpurua}}, \ and\ \bibinfo {author}
  {\bibfnamefont {B.}~\bibnamefont {Chichkov}},\ }\href@noop {} {\bibfield
  {journal} {\bibinfo  {journal} {ACS Photonics}\ }\textbf {\bibinfo {volume}
  {2}},\ \bibinfo {pages} {913} (\bibinfo {year} {2015})}\BibitemShut {NoStop}%
\bibitem [{\citenamefont {Campione}\ \emph {et~al.}(2016)\citenamefont
  {Campione}, \citenamefont {Liu}, \citenamefont {Basilio}, \citenamefont
  {Warne}, \citenamefont {Langston}, \citenamefont {Luk}, \citenamefont
  {Wendt}, \citenamefont {Reno}, \citenamefont {Keeler}, \citenamefont {Brener}
  \emph {et~al.}}]{campione2016broken}%
  \BibitemOpen
  \bibfield  {author} {\bibinfo {author} {\bibfnamefont {S.}~\bibnamefont
  {Campione}}, \bibinfo {author} {\bibfnamefont {S.}~\bibnamefont {Liu}},
  \bibinfo {author} {\bibfnamefont {L.~I.}\ \bibnamefont {Basilio}}, \bibinfo
  {author} {\bibfnamefont {L.~K.}\ \bibnamefont {Warne}}, \bibinfo {author}
  {\bibfnamefont {W.~L.}\ \bibnamefont {Langston}}, \bibinfo {author}
  {\bibfnamefont {T.~S.}\ \bibnamefont {Luk}}, \bibinfo {author} {\bibfnamefont
  {J.~R.}\ \bibnamefont {Wendt}}, \bibinfo {author} {\bibfnamefont {J.~L.}\
  \bibnamefont {Reno}}, \bibinfo {author} {\bibfnamefont {G.~A.}\ \bibnamefont
  {Keeler}}, \bibinfo {author} {\bibfnamefont {I.}~\bibnamefont {Brener}},
  \emph {et~al.},\ }\href@noop {} {\bibfield  {journal} {\bibinfo  {journal}
  {ACS Photonics}\ }\textbf {\bibinfo {volume} {3}},\ \bibinfo {pages} {2362}
  (\bibinfo {year} {2016})}\BibitemShut {NoStop}%
\bibitem [{\citenamefont {Shi}\ \emph {et~al.}(2012)\citenamefont {Shi},
  \citenamefont {Tuzer}, \citenamefont {Fenollosa},\ and\ \citenamefont
  {Meseguer}}]{ShiAdvMat2012}%
  \BibitemOpen
  \bibfield  {author} {\bibinfo {author} {\bibfnamefont {L.}~\bibnamefont
  {Shi}}, \bibinfo {author} {\bibfnamefont {T.}~\bibnamefont {Tuzer}}, \bibinfo
  {author} {\bibfnamefont {R.}~\bibnamefont {Fenollosa}}, \ and\ \bibinfo
  {author} {\bibfnamefont {F.}~\bibnamefont {Meseguer}},\ }\href@noop {}
  {\bibfield  {journal} {\bibinfo  {journal} {Adv. Mater.}\ }\textbf {\bibinfo
  {volume} {24}},\ \bibinfo {pages} {5934} (\bibinfo {year}
  {2012})}\BibitemShut {NoStop}%
\bibitem [{\citenamefont {van~de Haar}\ \emph {et~al.}(2016)\citenamefont
  {van~de Haar}, \citenamefont {van~de Groep}, \citenamefont {Brenny},\ and\
  \citenamefont {Polman}}]{Polman}%
  \BibitemOpen
  \bibfield  {author} {\bibinfo {author} {\bibfnamefont {M.}~\bibnamefont
  {van~de Haar}}, \bibinfo {author} {\bibfnamefont {J.}~\bibnamefont {van~de
  Groep}}, \bibinfo {author} {\bibfnamefont {B.}~\bibnamefont {Brenny}}, \ and\
  \bibinfo {author} {\bibfnamefont {A.}~\bibnamefont {Polman}},\ }\href@noop {}
  {\bibfield  {journal} {\bibinfo  {journal} {Opt. Express}\ }\textbf {\bibinfo
  {volume} {24}},\ \bibinfo {pages} {2047} (\bibinfo {year}
  {2016})}\BibitemShut {NoStop}%
\bibitem [{\citenamefont {Lee}\ \emph {et~al.}(2015)\citenamefont {Lee},
  \citenamefont {Song}, \citenamefont {Jeong}, \citenamefont {Kang},
  \citenamefont {Park},\ and\ \citenamefont {Seo}}]{Seo}%
  \BibitemOpen
  \bibfield  {author} {\bibinfo {author} {\bibfnamefont {E.-K.}\ \bibnamefont
  {Lee}}, \bibinfo {author} {\bibfnamefont {J.-H.}\ \bibnamefont {Song}},
  \bibinfo {author} {\bibfnamefont {K.-Y.}\ \bibnamefont {Jeong}}, \bibinfo
  {author} {\bibfnamefont {J.-H.}\ \bibnamefont {Kang}}, \bibinfo {author}
  {\bibfnamefont {H.-G.}\ \bibnamefont {Park}}, \ and\ \bibinfo {author}
  {\bibfnamefont {M.-K.}\ \bibnamefont {Seo}},\ }\href@noop {} {\bibfield
  {journal} {\bibinfo  {journal} {Sci. Rep.}\ }\textbf {\bibinfo {volume}
  {5}},\ \bibinfo {pages} {10400} (\bibinfo {year} {2015})}\BibitemShut
  {NoStop}%
\bibitem [{\citenamefont {Feng}\ \emph {et~al.}(2010)\citenamefont {Feng},
  \citenamefont {Li},\ and\ \citenamefont {Fan}}]{Feng}%
  \BibitemOpen
  \bibfield  {author} {\bibinfo {author} {\bibfnamefont {J.}~\bibnamefont
  {Feng}}, \bibinfo {author} {\bibfnamefont {Q.}~\bibnamefont {Li}}, \ and\
  \bibinfo {author} {\bibfnamefont {S.}~\bibnamefont {Fan}},\ }\href@noop {}
  {\bibfield  {journal} {\bibinfo  {journal} {Opt. Lett.}\ }\textbf {\bibinfo
  {volume} {35}},\ \bibinfo {pages} {3904} (\bibinfo {year}
  {2010})}\BibitemShut {NoStop}%
\bibitem [{\citenamefont {I.Staude}\ \emph {et~al.}(2013)\citenamefont
  {I.Staude}, \citenamefont {Miroshnichenko}, \citenamefont {M.Decker},
  \citenamefont {Fofang}, \citenamefont {S.Liu}, \citenamefont {E.Gonzales},
  \citenamefont {J.Dominguez}, \citenamefont {Luk}, \citenamefont {Neshev},
  \citenamefont {I.Brener},\ and\ \citenamefont
  {Y.Kivshar}}]{StaudeACSNano2013}%
  \BibitemOpen
  \bibfield  {author} {\bibinfo {author} {\bibnamefont {I.Staude}}, \bibinfo
  {author} {\bibfnamefont {A.}~\bibnamefont {Miroshnichenko}}, \bibinfo
  {author} {\bibnamefont {M.Decker}}, \bibinfo {author} {\bibfnamefont
  {N.}~\bibnamefont {Fofang}}, \bibinfo {author} {\bibnamefont {S.Liu}},
  \bibinfo {author} {\bibnamefont {E.Gonzales}}, \bibinfo {author}
  {\bibnamefont {J.Dominguez}}, \bibinfo {author} {\bibfnamefont
  {T.}~\bibnamefont {Luk}}, \bibinfo {author} {\bibfnamefont {D.}~\bibnamefont
  {Neshev}}, \bibinfo {author} {\bibnamefont {I.Brener}}, \ and\ \bibinfo
  {author} {\bibnamefont {Y.Kivshar}},\ }\href@noop {} {\bibfield  {journal}
  {\bibinfo  {journal} {ACS Nano}\ }\textbf {\bibinfo {volume} {7}},\ \bibinfo
  {pages} {7824} (\bibinfo {year} {2013})}\BibitemShut {NoStop}%
\bibitem [{\citenamefont {Spinelli}\ \emph {et~al.}(2012)\citenamefont
  {Spinelli}, \citenamefont {Verschuuren},\ and\ \citenamefont
  {Polman}}]{SpinelliNC2012}%
  \BibitemOpen
  \bibfield  {author} {\bibinfo {author} {\bibfnamefont {P.}~\bibnamefont
  {Spinelli}}, \bibinfo {author} {\bibfnamefont {M.}~\bibnamefont
  {Verschuuren}}, \ and\ \bibinfo {author} {\bibfnamefont {A.}~\bibnamefont
  {Polman}},\ }\href@noop {} {\bibfield  {journal} {\bibinfo  {journal} {Nat.
  Commun.}\ }\textbf {\bibinfo {volume} {3}},\ \bibinfo {pages} {692} (\bibinfo
  {year} {2012})}\BibitemShut {NoStop}%
\bibitem [{\citenamefont {Person}\ \emph {et~al.}(2013)\citenamefont {Person},
  \citenamefont {Jain}, \citenamefont {Lapin}, \citenamefont {Saenz},
  \citenamefont {Wicks},\ and\ \citenamefont {Novotny}}]{Novotny2013}%
  \BibitemOpen
  \bibfield  {author} {\bibinfo {author} {\bibfnamefont {S.}~\bibnamefont
  {Person}}, \bibinfo {author} {\bibfnamefont {M.}~\bibnamefont {Jain}},
  \bibinfo {author} {\bibfnamefont {Z.}~\bibnamefont {Lapin}}, \bibinfo
  {author} {\bibfnamefont {J.~J.}\ \bibnamefont {Saenz}}, \bibinfo {author}
  {\bibfnamefont {G.}~\bibnamefont {Wicks}}, \ and\ \bibinfo {author}
  {\bibfnamefont {L.}~\bibnamefont {Novotny}},\ }\href@noop {} {\bibfield
  {journal} {\bibinfo  {journal} {Nano Lett.}\ }\textbf {\bibinfo {volume}
  {13}},\ \bibinfo {pages} {1806} (\bibinfo {year} {2013})}\BibitemShut
  {NoStop}%
\bibitem [{\citenamefont {Liu}\ \emph {et~al.}(2013)\citenamefont {Liu},
  \citenamefont {Ihlefeld}, \citenamefont {Dominguez}, \citenamefont
  {Gonzales}, \citenamefont {Bower}, \citenamefont {Burckel}, \citenamefont
  {Sinclair},\ and\ \citenamefont {Brener}}]{Sinclair2013}%
  \BibitemOpen
  \bibfield  {author} {\bibinfo {author} {\bibfnamefont {S.}~\bibnamefont
  {Liu}}, \bibinfo {author} {\bibfnamefont {J.~F.}\ \bibnamefont {Ihlefeld}},
  \bibinfo {author} {\bibfnamefont {J.}~\bibnamefont {Dominguez}}, \bibinfo
  {author} {\bibfnamefont {E.~F.}\ \bibnamefont {Gonzales}}, \bibinfo {author}
  {\bibfnamefont {J.~E.}\ \bibnamefont {Bower}}, \bibinfo {author}
  {\bibfnamefont {D.~B.}\ \bibnamefont {Burckel}}, \bibinfo {author}
  {\bibfnamefont {M.~B.}\ \bibnamefont {Sinclair}}, \ and\ \bibinfo {author}
  {\bibfnamefont {I.}~\bibnamefont {Brener}},\ }\href@noop {} {\bibfield
  {journal} {\bibinfo  {journal} {Appl. Phys. Lett.}\ }\textbf {\bibinfo
  {volume} {102}},\ \bibinfo {pages} {161905} (\bibinfo {year}
  {2013})}\BibitemShut {NoStop}%
\bibitem [{\citenamefont {Spinelli}\ \emph {et~al.}(2013)\citenamefont
  {Spinelli}, \citenamefont {Macco}, \citenamefont {Verschuuren}, \citenamefont
  {Kessels},\ and\ \citenamefont {Polman}}]{Polman2013}%
  \BibitemOpen
  \bibfield  {author} {\bibinfo {author} {\bibfnamefont {P.}~\bibnamefont
  {Spinelli}}, \bibinfo {author} {\bibfnamefont {B.}~\bibnamefont {Macco}},
  \bibinfo {author} {\bibfnamefont {M.~A.}\ \bibnamefont {Verschuuren}},
  \bibinfo {author} {\bibfnamefont {W.}~\bibnamefont {Kessels}}, \ and\
  \bibinfo {author} {\bibfnamefont {A.}~\bibnamefont {Polman}},\ }\href@noop {}
  {\bibfield  {journal} {\bibinfo  {journal} {Appl. Phys. Lett.}\ }\textbf
  {\bibinfo {volume} {102}},\ \bibinfo {pages} {233902} (\bibinfo {year}
  {2013})}\BibitemShut {NoStop}%
\bibitem [{\citenamefont {Decker}\ and\ \citenamefont
  {Staude}(2016)}]{StaudeReview}%
  \BibitemOpen
  \bibfield  {author} {\bibinfo {author} {\bibfnamefont {M.}~\bibnamefont
  {Decker}}\ and\ \bibinfo {author} {\bibfnamefont {I.}~\bibnamefont
  {Staude}},\ }\href@noop {} {\bibfield  {journal} {\bibinfo  {journal} {J.
  Opt.}\ }\textbf {\bibinfo {volume} {18}},\ \bibinfo {pages} {103001}
  (\bibinfo {year} {2016})}\BibitemShut {NoStop}%
\bibitem [{\citenamefont {Abbarchi}\ \emph {et~al.}(2014)\citenamefont
  {Abbarchi}, \citenamefont {Naffouti}, \citenamefont {Vial}, \citenamefont
  {Benkouider}, \citenamefont {Lermusiaux}, \citenamefont {Favre},
  \citenamefont {Ronda}, \citenamefont {Bidault}, \citenamefont {Berbezier},\
  and\ \citenamefont {Bonod}}]{AbbarchiACSNano2014}%
  \BibitemOpen
  \bibfield  {author} {\bibinfo {author} {\bibfnamefont {M.}~\bibnamefont
  {Abbarchi}}, \bibinfo {author} {\bibfnamefont {M.}~\bibnamefont {Naffouti}},
  \bibinfo {author} {\bibfnamefont {B.}~\bibnamefont {Vial}}, \bibinfo {author}
  {\bibfnamefont {A.}~\bibnamefont {Benkouider}}, \bibinfo {author}
  {\bibfnamefont {L.}~\bibnamefont {Lermusiaux}}, \bibinfo {author}
  {\bibfnamefont {L.}~\bibnamefont {Favre}}, \bibinfo {author} {\bibfnamefont
  {A.}~\bibnamefont {Ronda}}, \bibinfo {author} {\bibfnamefont
  {S.}~\bibnamefont {Bidault}}, \bibinfo {author} {\bibfnamefont
  {I.}~\bibnamefont {Berbezier}}, \ and\ \bibinfo {author} {\bibfnamefont
  {N.}~\bibnamefont {Bonod}},\ }\href@noop {} {\bibfield  {journal} {\bibinfo
  {journal} {ACS Nano}\ }\textbf {\bibinfo {volume} {8}},\ \bibinfo {pages}
  {11181} (\bibinfo {year} {2014})}\BibitemShut {NoStop}%
\bibitem [{\citenamefont {Naffouti}\ \emph {et~al.}(2016)\citenamefont
  {Naffouti}, \citenamefont {David}, \citenamefont {Benkouider}, \citenamefont
  {Favre}, \citenamefont {Ronda}, \citenamefont {Berbezier}, \citenamefont
  {Bidault}, \citenamefont {Bonod},\ and\ \citenamefont
  {Abbarchi}}]{AbbarchiNanoscale}%
  \BibitemOpen
  \bibfield  {author} {\bibinfo {author} {\bibfnamefont {M.}~\bibnamefont
  {Naffouti}}, \bibinfo {author} {\bibfnamefont {T.}~\bibnamefont {David}},
  \bibinfo {author} {\bibfnamefont {A.}~\bibnamefont {Benkouider}}, \bibinfo
  {author} {\bibfnamefont {L.}~\bibnamefont {Favre}}, \bibinfo {author}
  {\bibfnamefont {A.}~\bibnamefont {Ronda}}, \bibinfo {author} {\bibfnamefont
  {I.}~\bibnamefont {Berbezier}}, \bibinfo {author} {\bibfnamefont
  {S.}~\bibnamefont {Bidault}}, \bibinfo {author} {\bibfnamefont
  {N.}~\bibnamefont {Bonod}}, \ and\ \bibinfo {author} {\bibfnamefont
  {M.}~\bibnamefont {Abbarchi}},\ }\href@noop {} {\bibfield  {journal}
  {\bibinfo  {journal} {Nanoscale}\ }\textbf {\bibinfo {volume} {8}},\ \bibinfo
  {pages} {2844} (\bibinfo {year} {2016})}\BibitemShut {NoStop}%
\bibitem [{\citenamefont {Weidmann}\ and\ \citenamefont
  {Anderson}(1971)}]{Anderson}%
  \BibitemOpen
  \bibfield  {author} {\bibinfo {author} {\bibfnamefont {E.}~\bibnamefont
  {Weidmann}}\ and\ \bibinfo {author} {\bibfnamefont {J.}~\bibnamefont
  {Anderson}},\ }\href@noop {} {\bibfield  {journal} {\bibinfo  {journal} {Thin
  Solid Films}\ }\textbf {\bibinfo {volume} {7}},\ \bibinfo {pages} {265}
  (\bibinfo {year} {1971})}\BibitemShut {NoStop}%
\bibitem [{\citenamefont {Zhang}\ \emph {et~al.}(2009)\citenamefont {Zhang},
  \citenamefont {Yang}, \citenamefont {Rugheimer}, \citenamefont {Roberts},
  \citenamefont {Savage}, \citenamefont {Liu},\ and\ \citenamefont
  {Lagally}}]{Lagally}%
  \BibitemOpen
  \bibfield  {author} {\bibinfo {author} {\bibfnamefont {P.}~\bibnamefont
  {Zhang}}, \bibinfo {author} {\bibfnamefont {B.}~\bibnamefont {Yang}},
  \bibinfo {author} {\bibfnamefont {P.}~\bibnamefont {Rugheimer}}, \bibinfo
  {author} {\bibfnamefont {M.}~\bibnamefont {Roberts}}, \bibinfo {author}
  {\bibfnamefont {D.}~\bibnamefont {Savage}}, \bibinfo {author} {\bibfnamefont
  {F.}~\bibnamefont {Liu}}, \ and\ \bibinfo {author} {\bibfnamefont
  {M.}~\bibnamefont {Lagally}},\ }\href@noop {} {\bibfield  {journal} {\bibinfo
   {journal} {J. Phys. D: Appl. Phys.}\ }\textbf {\bibinfo {volume} {42}},\
  \bibinfo {pages} {175309} (\bibinfo {year} {2009})}\BibitemShut {NoStop}%
\bibitem [{\citenamefont {Proust}\ \emph {et~al.}(2015)\citenamefont {Proust},
  \citenamefont {Bedu}, \citenamefont {Chenot}, \citenamefont {Soumahoro},
  \citenamefont {Ozerov}, \citenamefont {Gallas}, \citenamefont {Abdeddaim},\
  and\ \citenamefont {Bonod}}]{Proust}%
  \BibitemOpen
  \bibfield  {author} {\bibinfo {author} {\bibfnamefont {J.}~\bibnamefont
  {Proust}}, \bibinfo {author} {\bibfnamefont {F.}~\bibnamefont {Bedu}},
  \bibinfo {author} {\bibfnamefont {S.}~\bibnamefont {Chenot}}, \bibinfo
  {author} {\bibfnamefont {I.}~\bibnamefont {Soumahoro}}, \bibinfo {author}
  {\bibfnamefont {I.}~\bibnamefont {Ozerov}}, \bibinfo {author} {\bibfnamefont
  {B.}~\bibnamefont {Gallas}}, \bibinfo {author} {\bibfnamefont
  {R.}~\bibnamefont {Abdeddaim}}, \ and\ \bibinfo {author} {\bibfnamefont
  {N.}~\bibnamefont {Bonod}},\ }\href@noop {} {\bibfield  {journal} {\bibinfo
  {journal} {Adv. Opt. Mater.}\ }\textbf {\bibinfo {volume} {3}},\ \bibinfo
  {pages} {1280} (\bibinfo {year} {2015})}\BibitemShut {NoStop}%
\bibitem [{\citenamefont {Shi}\ \emph {et~al.}(2013)\citenamefont {Shi},
  \citenamefont {Harris}, \citenamefont {Fenollosa}, \citenamefont {Rodriguez},
  \citenamefont {Lu}, \citenamefont {Korgel},\ and\ \citenamefont
  {Meseguer}}]{ShiNC2013}%
  \BibitemOpen
  \bibfield  {author} {\bibinfo {author} {\bibfnamefont {L.}~\bibnamefont
  {Shi}}, \bibinfo {author} {\bibfnamefont {J.}~\bibnamefont {Harris}},
  \bibinfo {author} {\bibfnamefont {R.}~\bibnamefont {Fenollosa}}, \bibinfo
  {author} {\bibfnamefont {I.}~\bibnamefont {Rodriguez}}, \bibinfo {author}
  {\bibfnamefont {X.}~\bibnamefont {Lu}}, \bibinfo {author} {\bibfnamefont
  {B.}~\bibnamefont {Korgel}}, \ and\ \bibinfo {author} {\bibfnamefont
  {F.}~\bibnamefont {Meseguer}},\ }\href@noop {} {\bibfield  {journal}
  {\bibinfo  {journal} {Nat. Commun.}\ }\textbf {\bibinfo {volume} {4}},\
  \bibinfo {pages} {1904} (\bibinfo {year} {2013})}\BibitemShut {NoStop}%
\bibitem [{\citenamefont {Kim}\ \emph {et~al.}(2014)\citenamefont {Kim},
  \citenamefont {Thomann}, \citenamefont {Park}, \citenamefont {Kang},
  \citenamefont {Vasudev},\ and\ \citenamefont {Brongersma}}]{Brongersma}%
  \BibitemOpen
  \bibfield  {author} {\bibinfo {author} {\bibfnamefont {S.~J.}\ \bibnamefont
  {Kim}}, \bibinfo {author} {\bibfnamefont {I.}~\bibnamefont {Thomann}},
  \bibinfo {author} {\bibfnamefont {J.}~\bibnamefont {Park}}, \bibinfo {author}
  {\bibfnamefont {J.-H.}\ \bibnamefont {Kang}}, \bibinfo {author}
  {\bibfnamefont {A.~P.}\ \bibnamefont {Vasudev}}, \ and\ \bibinfo {author}
  {\bibfnamefont {M.~L.}\ \bibnamefont {Brongersma}},\ }\href@noop {}
  {\bibfield  {journal} {\bibinfo  {journal} {Nano Lett.}\ }\textbf {\bibinfo
  {volume} {14}},\ \bibinfo {pages} {1446} (\bibinfo {year}
  {2014})}\BibitemShut {NoStop}%
\bibitem [{\citenamefont {Lewi}\ \emph {et~al.}(2015)\citenamefont {Lewi},
  \citenamefont {Iyer}, \citenamefont {Butakov}, \citenamefont {Mikhailovsky},\
  and\ \citenamefont {Schuller}}]{Lewi2015}%
  \BibitemOpen
  \bibfield  {author} {\bibinfo {author} {\bibfnamefont {T.}~\bibnamefont
  {Lewi}}, \bibinfo {author} {\bibfnamefont {P.~P.}\ \bibnamefont {Iyer}},
  \bibinfo {author} {\bibfnamefont {N.~A.}\ \bibnamefont {Butakov}}, \bibinfo
  {author} {\bibfnamefont {A.~A.}\ \bibnamefont {Mikhailovsky}}, \ and\
  \bibinfo {author} {\bibfnamefont {J.~A.}\ \bibnamefont {Schuller}},\ }\href
  {\doibase 10.1021/acs.nanolett.5b03679} {\bibfield  {journal} {\bibinfo
  {journal} {Nano Lett.}\ }\textbf {\bibinfo {volume} {15}},\ \bibinfo {pages}
  {8188} (\bibinfo {year} {2015})}\BibitemShut {NoStop}%
\bibitem [{\citenamefont {Okamoto}\ \emph {et~al.}(2014)\citenamefont
  {Okamoto}, \citenamefont {Inaba}, \citenamefont {Iida}, \citenamefont
  {Ishihara}, \citenamefont {Ichikawa},\ and\ \citenamefont
  {Ashida}}]{Okamoto}%
  \BibitemOpen
  \bibfield  {author} {\bibinfo {author} {\bibfnamefont {S.}~\bibnamefont
  {Okamoto}}, \bibinfo {author} {\bibfnamefont {K.}~\bibnamefont {Inaba}},
  \bibinfo {author} {\bibfnamefont {T.}~\bibnamefont {Iida}}, \bibinfo {author}
  {\bibfnamefont {H.}~\bibnamefont {Ishihara}}, \bibinfo {author}
  {\bibfnamefont {S.}~\bibnamefont {Ichikawa}}, \ and\ \bibinfo {author}
  {\bibfnamefont {M.}~\bibnamefont {Ashida}},\ }\href@noop {} {\bibfield
  {journal} {\bibinfo  {journal} {Sci. Rep.}\ }\textbf {\bibinfo {volume}
  {4}},\ \bibinfo {pages} {5186} (\bibinfo {year} {2014})}\BibitemShut
  {NoStop}%
\bibitem [{\citenamefont {Zywietz}\ \emph
  {et~al.}(2014{\natexlab{b}})\citenamefont {Zywietz}, \citenamefont
  {Reinhardt}, \citenamefont {Evlyukhin}, \citenamefont {Birr},\ and\
  \citenamefont {Chichkov}}]{ZywietzAPA2014}%
  \BibitemOpen
  \bibfield  {author} {\bibinfo {author} {\bibfnamefont {U.}~\bibnamefont
  {Zywietz}}, \bibinfo {author} {\bibfnamefont {C.}~\bibnamefont {Reinhardt}},
  \bibinfo {author} {\bibfnamefont {A.}~\bibnamefont {Evlyukhin}}, \bibinfo
  {author} {\bibfnamefont {T.}~\bibnamefont {Birr}}, \ and\ \bibinfo {author}
  {\bibfnamefont {B.}~\bibnamefont {Chichkov}},\ }\href@noop {} {\bibfield
  {journal} {\bibinfo  {journal} {Appl. Phys. A}\ }\textbf {\bibinfo {volume}
  {114}},\ \bibinfo {pages} {45} (\bibinfo {year}
  {2014}{\natexlab{b}})}\BibitemShut {NoStop}%
\bibitem [{\citenamefont {Fan}\ and\ \citenamefont {Chu}(2010)}]{Chu}%
  \BibitemOpen
  \bibfield  {author} {\bibinfo {author} {\bibfnamefont {J.}~\bibnamefont
  {Fan}}\ and\ \bibinfo {author} {\bibfnamefont {P.}~\bibnamefont {Chu}},\
  }\href@noop {} {\bibfield  {journal} {\bibinfo  {journal} {Small}\ }\textbf
  {\bibinfo {volume} {6}},\ \bibinfo {pages} {2080} (\bibinfo {year}
  {2010})}\BibitemShut {NoStop}%
\bibitem [{\citenamefont {Tamarov}\ \emph {et~al.}(2014)\citenamefont
  {Tamarov}, \citenamefont {Osminkina}, \citenamefont {Zinovyev}, \citenamefont
  {Maximova}, \citenamefont {Kargina}, \citenamefont {Gongalsky}, \citenamefont
  {Ryabchikov}, \citenamefont {A.Al-Kattan}, \citenamefont {Sviridov},
  \citenamefont {Sentis}, \citenamefont {Ivanov}, \citenamefont {Nikiforov},
  \citenamefont {Kabashin},\ and\ \citenamefont {Timoshenko}}]{Timoshenko}%
  \BibitemOpen
  \bibfield  {author} {\bibinfo {author} {\bibfnamefont {K.}~\bibnamefont
  {Tamarov}}, \bibinfo {author} {\bibfnamefont {L.}~\bibnamefont {Osminkina}},
  \bibinfo {author} {\bibfnamefont {S.}~\bibnamefont {Zinovyev}}, \bibinfo
  {author} {\bibfnamefont {K.}~\bibnamefont {Maximova}}, \bibinfo {author}
  {\bibfnamefont {J.}~\bibnamefont {Kargina}}, \bibinfo {author} {\bibfnamefont
  {M.}~\bibnamefont {Gongalsky}}, \bibinfo {author} {\bibfnamefont
  {Y.}~\bibnamefont {Ryabchikov}}, \bibinfo {author} {\bibnamefont
  {A.Al-Kattan}}, \bibinfo {author} {\bibfnamefont {A.}~\bibnamefont
  {Sviridov}}, \bibinfo {author} {\bibfnamefont {M.}~\bibnamefont {Sentis}},
  \bibinfo {author} {\bibfnamefont {A.}~\bibnamefont {Ivanov}}, \bibinfo
  {author} {\bibfnamefont {V.}~\bibnamefont {Nikiforov}}, \bibinfo {author}
  {\bibfnamefont {A.}~\bibnamefont {Kabashin}}, \ and\ \bibinfo {author}
  {\bibfnamefont {V.}~\bibnamefont {Timoshenko}},\ }\href@noop {} {\bibfield
  {journal} {\bibinfo  {journal} {Sci. Rep.}\ }\textbf {\bibinfo {volume}
  {4}},\ \bibinfo {pages} {7034} (\bibinfo {year} {2014})}\BibitemShut
  {NoStop}%
\bibitem [{\citenamefont {Blandin}\ \emph {et~al.}(2013)\citenamefont
  {Blandin}, \citenamefont {Maximova}, \citenamefont {Gongalsky}, \citenamefont
  {Sanchez-Royo}, \citenamefont {Chirvony}, \citenamefont {Sentis},
  \citenamefont {Timoshenko},\ and\ \citenamefont {Kabashin}}]{KabashinJMCB}%
  \BibitemOpen
  \bibfield  {author} {\bibinfo {author} {\bibfnamefont {P.}~\bibnamefont
  {Blandin}}, \bibinfo {author} {\bibfnamefont {K.}~\bibnamefont {Maximova}},
  \bibinfo {author} {\bibfnamefont {M.}~\bibnamefont {Gongalsky}}, \bibinfo
  {author} {\bibfnamefont {J.}~\bibnamefont {Sanchez-Royo}}, \bibinfo {author}
  {\bibfnamefont {V.}~\bibnamefont {Chirvony}}, \bibinfo {author}
  {\bibfnamefont {M.}~\bibnamefont {Sentis}}, \bibinfo {author} {\bibfnamefont
  {V.}~\bibnamefont {Timoshenko}}, \ and\ \bibinfo {author} {\bibfnamefont
  {A.}~\bibnamefont {Kabashin}},\ }\href@noop {} {\bibfield  {journal}
  {\bibinfo  {journal} {J. Mater. Chem. B}\ }\textbf {\bibinfo {volume} {1}},\
  \bibinfo {pages} {2489} (\bibinfo {year} {2013})}\BibitemShut {NoStop}%
\bibitem [{\citenamefont {Gu}\ \emph {et~al.}(2013)\citenamefont {Gu},
  \citenamefont {Hall}, \citenamefont {Qin}, \citenamefont {Anglin},
  \citenamefont {Joo}, \citenamefont {Mooney}, \citenamefont {Howell},\ and\
  \citenamefont {Sailor}}]{Sailor}%
  \BibitemOpen
  \bibfield  {author} {\bibinfo {author} {\bibfnamefont {L.}~\bibnamefont
  {Gu}}, \bibinfo {author} {\bibfnamefont {D.}~\bibnamefont {Hall}}, \bibinfo
  {author} {\bibfnamefont {Z.}~\bibnamefont {Qin}}, \bibinfo {author}
  {\bibfnamefont {E.}~\bibnamefont {Anglin}}, \bibinfo {author} {\bibfnamefont
  {J.}~\bibnamefont {Joo}}, \bibinfo {author} {\bibfnamefont {D.}~\bibnamefont
  {Mooney}}, \bibinfo {author} {\bibfnamefont {S.}~\bibnamefont {Howell}}, \
  and\ \bibinfo {author} {\bibfnamefont {M.}~\bibnamefont {Sailor}},\
  }\href@noop {} {\bibfield  {journal} {\bibinfo  {journal} {Nat. Commun.}\
  }\textbf {\bibinfo {volume} {4}},\ \bibinfo {pages} {2326} (\bibinfo {year}
  {2013})}\BibitemShut {NoStop}%
\bibitem [{\citenamefont {Kormer}\ \emph {et~al.}(2012)\citenamefont {Kormer},
  \citenamefont {Butz}, \citenamefont {Spiecker},\ and\ \citenamefont
  {Peukert}}]{Peukert}%
  \BibitemOpen
  \bibfield  {author} {\bibinfo {author} {\bibfnamefont {R.}~\bibnamefont
  {Kormer}}, \bibinfo {author} {\bibfnamefont {B.}~\bibnamefont {Butz}},
  \bibinfo {author} {\bibfnamefont {E.}~\bibnamefont {Spiecker}}, \ and\
  \bibinfo {author} {\bibfnamefont {W.}~\bibnamefont {Peukert}},\ }\href@noop
  {} {\bibfield  {journal} {\bibinfo  {journal} {Cryst. Growth Des.}\ }\textbf
  {\bibinfo {volume} {12}},\ \bibinfo {pages} {1330} (\bibinfo {year}
  {2012})}\BibitemShut {NoStop}%
\bibitem [{\citenamefont {Devlin}\ \emph {et~al.}(2016)\citenamefont {Devlin},
  \citenamefont {Khorasaninejad}, \citenamefont {Chen}, \citenamefont {Oh},\
  and\ \citenamefont {Capasso}}]{devlin2016broadband}%
  \BibitemOpen
  \bibfield  {author} {\bibinfo {author} {\bibfnamefont {R.~C.}\ \bibnamefont
  {Devlin}}, \bibinfo {author} {\bibfnamefont {M.}~\bibnamefont
  {Khorasaninejad}}, \bibinfo {author} {\bibfnamefont {W.~T.}\ \bibnamefont
  {Chen}}, \bibinfo {author} {\bibfnamefont {J.}~\bibnamefont {Oh}}, \ and\
  \bibinfo {author} {\bibfnamefont {F.}~\bibnamefont {Capasso}},\ }\href@noop
  {} {\bibfield  {journal} {\bibinfo  {journal} {Proceedings of the National
  Academy of Sciences}\ }\textbf {\bibinfo {volume} {113}},\ \bibinfo {pages}
  {10473} (\bibinfo {year} {2016})}\BibitemShut {NoStop}%
\bibitem [{\citenamefont {Genevet}\ \emph {et~al.}(2017)\citenamefont
  {Genevet}, \citenamefont {Capasso}, \citenamefont {Aieta}, \citenamefont
  {Khorasaninejad},\ and\ \citenamefont {Devlin}}]{genevet2017recent}%
  \BibitemOpen
  \bibfield  {author} {\bibinfo {author} {\bibfnamefont {P.}~\bibnamefont
  {Genevet}}, \bibinfo {author} {\bibfnamefont {F.}~\bibnamefont {Capasso}},
  \bibinfo {author} {\bibfnamefont {F.}~\bibnamefont {Aieta}}, \bibinfo
  {author} {\bibfnamefont {M.}~\bibnamefont {Khorasaninejad}}, \ and\ \bibinfo
  {author} {\bibfnamefont {R.}~\bibnamefont {Devlin}},\ }\href@noop {}
  {\bibfield  {journal} {\bibinfo  {journal} {Optica}\ }\textbf {\bibinfo
  {volume} {4}},\ \bibinfo {pages} {139} (\bibinfo {year} {2017})}\BibitemShut
  {NoStop}%
\bibitem [{\citenamefont {O'Mara}\ \emph {et~al.}(1990)\citenamefont {O'Mara},
  \citenamefont {Herring},\ and\ \citenamefont {Hunt}}]{o1990handbook}%
  \BibitemOpen
  \bibfield  {author} {\bibinfo {author} {\bibfnamefont {W.}~\bibnamefont
  {O'Mara}}, \bibinfo {author} {\bibfnamefont {R.~B.}\ \bibnamefont {Herring}},
  \ and\ \bibinfo {author} {\bibfnamefont {L.~P.}\ \bibnamefont {Hunt}},\
  }\href@noop {} {\emph {\bibinfo {title} {Handbook of semiconductor silicon
  technology}}}\ (\bibinfo  {publisher} {Noyes Publications},\ \bibinfo {year}
  {1990})\BibitemShut {NoStop}%
\bibitem [{\citenamefont {Ye}\ and\ \citenamefont
  {Thompson}(2011)}]{ye2011templated}%
  \BibitemOpen
  \bibfield  {author} {\bibinfo {author} {\bibfnamefont {J.}~\bibnamefont
  {Ye}}\ and\ \bibinfo {author} {\bibfnamefont {C.}~\bibnamefont {Thompson}},\
  }\href@noop {} {\bibfield  {journal} {\bibinfo  {journal} {Adv. Mater.}\
  }\textbf {\bibinfo {volume} {23}},\ \bibinfo {pages} {1567} (\bibinfo {year}
  {2011})}\BibitemShut {NoStop}%
\bibitem [{\citenamefont {Thompson}(2012)}]{thompson2012solid}%
  \BibitemOpen
  \bibfield  {author} {\bibinfo {author} {\bibfnamefont {C.~V.}\ \bibnamefont
  {Thompson}},\ }\href@noop {} {\bibfield  {journal} {\bibinfo  {journal}
  {Annu. Rev. Mater. Res.}\ }\textbf {\bibinfo {volume} {42}},\ \bibinfo
  {pages} {399} (\bibinfo {year} {2012})}\BibitemShut {NoStop}%
\bibitem [{\citenamefont {Yan}\ \emph {et~al.}(2015)\citenamefont {Yan},
  \citenamefont {Liu}, \citenamefont {Lin}, \citenamefont {Wang}, \citenamefont
  {Chen}, \citenamefont {Wang},\ and\ \citenamefont {Yang}}]{Yang}%
  \BibitemOpen
  \bibfield  {author} {\bibinfo {author} {\bibfnamefont {J.}~\bibnamefont
  {Yan}}, \bibinfo {author} {\bibfnamefont {P.}~\bibnamefont {Liu}}, \bibinfo
  {author} {\bibfnamefont {Z.}~\bibnamefont {Lin}}, \bibinfo {author}
  {\bibfnamefont {H.}~\bibnamefont {Wang}}, \bibinfo {author} {\bibfnamefont
  {H.}~\bibnamefont {Chen}}, \bibinfo {author} {\bibfnamefont {C.}~\bibnamefont
  {Wang}}, \ and\ \bibinfo {author} {\bibfnamefont {G.}~\bibnamefont {Yang}},\
  }\href@noop {} {\bibfield  {journal} {\bibinfo  {journal} {Nat. Commun.}\
  }\textbf {\bibinfo {volume} {6}},\ \bibinfo {pages} {7042} (\bibinfo {year}
  {2015})}\BibitemShut {NoStop}%
\bibitem [{\citenamefont {Dmitriev}\ \emph {et~al.}(2015)\citenamefont
  {Dmitriev}, \citenamefont {Makarov}, \citenamefont {Milichko}, \citenamefont
  {Mukhin}, \citenamefont {Gudovskikh}, \citenamefont {Sitnikova},
  \citenamefont {Samusev}, \citenamefont {Krasnok},\ and\ \citenamefont
  {Belov}}]{DmitrievNanoscale2015}%
  \BibitemOpen
  \bibfield  {author} {\bibinfo {author} {\bibfnamefont {P.}~\bibnamefont
  {Dmitriev}}, \bibinfo {author} {\bibfnamefont {S.}~\bibnamefont {Makarov}},
  \bibinfo {author} {\bibfnamefont {V.}~\bibnamefont {Milichko}}, \bibinfo
  {author} {\bibfnamefont {I.}~\bibnamefont {Mukhin}}, \bibinfo {author}
  {\bibfnamefont {A.}~\bibnamefont {Gudovskikh}}, \bibinfo {author}
  {\bibfnamefont {A.}~\bibnamefont {Sitnikova}}, \bibinfo {author}
  {\bibfnamefont {A.}~\bibnamefont {Samusev}}, \bibinfo {author} {\bibfnamefont
  {A.}~\bibnamefont {Krasnok}}, \ and\ \bibinfo {author} {\bibfnamefont
  {P.}~\bibnamefont {Belov}},\ }\href@noop {} {\bibfield  {journal} {\bibinfo
  {journal} {Nanoscale}\ }\textbf {\bibinfo {volume} {8}},\ \bibinfo {pages}
  {5043} (\bibinfo {year} {2015})}\BibitemShut {NoStop}%
\bibitem [{\citenamefont {Yoo}\ \emph {et~al.}(2015)\citenamefont {Yoo},
  \citenamefont {In}, \citenamefont {Zheng}, \citenamefont {Sakellari},
  \citenamefont {Raman}, \citenamefont {Matthews}, \citenamefont {Elhadj},\
  and\ \citenamefont {Grigoropoulos}}]{CostasNanotech2015}%
  \BibitemOpen
  \bibfield  {author} {\bibinfo {author} {\bibfnamefont {J.-H.}\ \bibnamefont
  {Yoo}}, \bibinfo {author} {\bibfnamefont {J.}~\bibnamefont {In}}, \bibinfo
  {author} {\bibfnamefont {C.}~\bibnamefont {Zheng}}, \bibinfo {author}
  {\bibfnamefont {I.}~\bibnamefont {Sakellari}}, \bibinfo {author}
  {\bibfnamefont {R.}~\bibnamefont {Raman}}, \bibinfo {author} {\bibfnamefont
  {M.}~\bibnamefont {Matthews}}, \bibinfo {author} {\bibfnamefont
  {S.}~\bibnamefont {Elhadj}}, \ and\ \bibinfo {author} {\bibfnamefont
  {C.}~\bibnamefont {Grigoropoulos}},\ }\href@noop {} {\bibfield  {journal}
  {\bibinfo  {journal} {Nanotechnology}\ }\textbf {\bibinfo {volume} {26}},\
  \bibinfo {pages} {165303} (\bibinfo {year} {2015})}\BibitemShut {NoStop}%
\bibitem [{\citenamefont {Bohandy}\ \emph {et~al.}(1988)\citenamefont
  {Bohandy}, \citenamefont {Kim}, \citenamefont {Adrian},\ and\ \citenamefont
  {Jette}}]{bohandy1988metal}%
  \BibitemOpen
  \bibfield  {author} {\bibinfo {author} {\bibfnamefont {J.}~\bibnamefont
  {Bohandy}}, \bibinfo {author} {\bibfnamefont {B.}~\bibnamefont {Kim}},
  \bibinfo {author} {\bibfnamefont {F.}~\bibnamefont {Adrian}}, \ and\ \bibinfo
  {author} {\bibfnamefont {A.}~\bibnamefont {Jette}},\ }\href@noop {}
  {\bibfield  {journal} {\bibinfo  {journal} {J. Appl. Phys.}\ }\textbf
  {\bibinfo {volume} {63}},\ \bibinfo {pages} {1158} (\bibinfo {year}
  {1988})}\BibitemShut {NoStop}%
\bibitem [{\citenamefont {Gibson}\ \emph {et~al.}(2015)\citenamefont {Gibson},
  \citenamefont {Schoop}, \citenamefont {Muechler}, \citenamefont {Xie},
  \citenamefont {Hirschberger}, \citenamefont {Ong}, \citenamefont {Car},\ and\
  \citenamefont {Cava}}]{Cava}%
  \BibitemOpen
  \bibfield  {author} {\bibinfo {author} {\bibfnamefont {Q.~D.}\ \bibnamefont
  {Gibson}}, \bibinfo {author} {\bibfnamefont {L.~M.}\ \bibnamefont {Schoop}},
  \bibinfo {author} {\bibfnamefont {L.}~\bibnamefont {Muechler}}, \bibinfo
  {author} {\bibfnamefont {L.~S.}\ \bibnamefont {Xie}}, \bibinfo {author}
  {\bibfnamefont {M.}~\bibnamefont {Hirschberger}}, \bibinfo {author}
  {\bibfnamefont {N.~P.}\ \bibnamefont {Ong}}, \bibinfo {author} {\bibfnamefont
  {R.}~\bibnamefont {Car}}, \ and\ \bibinfo {author} {\bibfnamefont {R.~J.}\
  \bibnamefont {Cava}},\ }\href@noop {} {\bibfield  {journal} {\bibinfo
  {journal} {Phys. Rev. B}\ }\textbf {\bibinfo {volume} {91}},\ \bibinfo
  {pages} {205128} (\bibinfo {year} {2015})}\BibitemShut {NoStop}%
\bibitem [{\citenamefont {Wehling}\ \emph {et~al.}(2014)\citenamefont
  {Wehling}, \citenamefont {Black-Schaffer},\ and\ \citenamefont
  {Balatsky}}]{Dirac}%
  \BibitemOpen
  \bibfield  {author} {\bibinfo {author} {\bibfnamefont {T.}~\bibnamefont
  {Wehling}}, \bibinfo {author} {\bibfnamefont {A.}~\bibnamefont
  {Black-Schaffer}}, \ and\ \bibinfo {author} {\bibfnamefont {A.}~\bibnamefont
  {Balatsky}},\ }\href@noop {} {\bibfield  {journal} {\bibinfo  {journal} {Adv.
  Phys.}\ }\textbf {\bibinfo {volume} {63}},\ \bibinfo {pages} {1} (\bibinfo
  {year} {2014})}\BibitemShut {NoStop}%
\bibitem [{\citenamefont {Weng}\ \emph {et~al.}(2016)\citenamefont {Weng},
  \citenamefont {Dai},\ and\ \citenamefont {Fang}}]{Topological}%
  \BibitemOpen
  \bibfield  {author} {\bibinfo {author} {\bibfnamefont {H.}~\bibnamefont
  {Weng}}, \bibinfo {author} {\bibfnamefont {X.}~\bibnamefont {Dai}}, \ and\
  \bibinfo {author} {\bibfnamefont {Z.}~\bibnamefont {Fang}},\ }\href@noop {}
  {\bibfield  {journal} {\bibinfo  {journal} {J. Phys. Condens. Matter}\
  }\textbf {\bibinfo {volume} {28}},\ \bibinfo {pages} {303001} (\bibinfo
  {year} {2016})}\BibitemShut {NoStop}%
\end{thebibliography}%

\end{document}